\documentclass[10pt,twocolumn,letterpaper]{article}
\usepackage{cvpr}

\usepackage[pagebackref,breaklinks,colorlinks]{hyperref}
\usepackage{times}
\usepackage{epsfig}
\usepackage{graphicx}
\usepackage{amsmath}
\usepackage{amssymb}
\usepackage{booktabs}

\usepackage{amsmath}
\usepackage{amssymb}
\usepackage{mathtools}
\usepackage{amsthm}
\usepackage{booktabs}
\usepackage{enumerate}
\usepackage{setspace}
\usepackage{amsthm}
\newcommand{\myparatight}[1]{\smallskip\noindent{\bf {#1}:}~}
\newcommand{\argmax}{\operatornamewithlimits{argmax}}

\newcommand{\lnorm}[1]{\ensuremath{\left\Vert#1\right\Vert}}
\newcommand{\name}{\text{PointCert}}
\newcommand{\pcc}{g}
\newcommand{\baseclf}{f}
\newcommand{\ensclf}{h}
\newcommand{\pointcloud}{P}
\usepackage[capitalize]{cleveref}
\crefname{section}{Sec.}{Secs.}
\Crefname{section}{Section}{Sections}
\Crefname{table}{Table}{Tables}
\crefname{table}{Tab.}{Tabs.}
\theoremstyle{plain}
\newtheorem{theorem}{Theorem}[section]

\theoremstyle{definition}

\theoremstyle{remark}

\def\cvprPaperID{5764} 
\def\confName{CVPR}
\def\confYear{2023}
\begin{document}

\title{PointCert: Point Cloud Classification with Deterministic \\Certified Robustness Guarantees}

\author{
   {Jinghuai Zhang}${^{1}}$ \quad
   {Jinyuan Jia}${^{2}}$ \quad 
   {Hongbin Liu}${^{1}}$ \quad
   {Neil Zhenqiang Gong}${^{1}}$\\
   {Duke University}${^1}$\qquad
   {UIUC}${^2}$ \qquad \\
{\tt\small \{jinghuai.zhang, hongbin.liu, neil.gong\}@duke.edu,} \\
{\tt\small \{jinyuan\}@illinois.edu,}
}

\maketitle

\begin{abstract}
Point cloud classification is an essential component in many security-critical applications such as autonomous driving and augmented reality. However, point cloud classifiers are vulnerable to adversarially perturbed point clouds. Existing certified defenses against adversarial point clouds suffer from a key limitation: their certified robustness guarantees are {probabilistic}, i.e., they produce an incorrect certified robustness guarantee with some probability. In this work, we propose a general framework, namely {\name}, that can transform an arbitrary point cloud classifier to be certifiably robust against adversarial point clouds with deterministic guarantees. {\name} certifiably predicts the same label for a point cloud when the number of arbitrarily added, deleted, and/or modified points is less than a threshold. Moreover, we propose multiple methods to optimize the certified robustness guarantees of {\name} in three application scenarios. We systematically evaluate {\name} on ModelNet and ScanObjectNN benchmark datasets. Our results show that {\name} substantially outperforms state-of-the-art certified defenses even though their robustness guarantees are probabilistic. 
\end{abstract}

\section{Introduction}
Point cloud classification \cite{qi2017pointnet,wang2019dynamic,qiu2021geometric,hamdi2021mvtn,zhao2021point,xiang2021walk} has many safety-critical applications, including but not limited to, autonomous driving and augmented reality. However, various studies~\cite{xiang2019generating, zhao2020isometry, kim2021minimal,liu2020adversarial,wicker2019robustness, zheng2019pointcloud,yang2019adversarial,hamdi2020advpc, shen2021interpreting} showed that point cloud classification is vulnerable to \emph{adversarial point clouds}. In particular, an attacker can carefully add, delete, and/or modify a small number of points in a point cloud to make it misclassified by a point cloud classifier. 

Existing defenses against adversarial point clouds  can be categorized into \emph{empirical defenses}~\cite{zhou2019dup,dong2020self,liu2019extending,yang2019adversarial,wu2020if,zhang2022pointcutmix,sun2021adversarially} and \emph{certified defenses}~\cite{cohen2019certified,liu2021pointguard, denipitiyage2021provable}. The key limitation of empirical defenses is that they cannot provide formal guarantees, and thus are often broken by advanced, adaptive attacks~\cite{sun2020adversarial}. Therefore, we focus on certified defenses in this work. Randomized smoothing~\cite{cohen2019certified} and PointGuard~\cite{liu2021pointguard} are two state-of-the-art certified defenses against adversarial point clouds. In particular, randomized smoothing adds random noise (e.g., Gaussian noise) to a point cloud, while PointGuard randomly subsamples a point cloud. 
Due to the randomness, their certified robustness guarantees are \emph{probabilistic}, i.e., they produce incorrect robustness guarantees   with some probability (called \emph{error probability}). For instance, when the error probability is 0.001, they produce incorrect robustness guarantees for 1 out of 1,000 point-cloud classifications on average. Such probabilistic guarantees are insufficient for security-critical applications that frequently classify point clouds. 

In this work, we propose {\name}, the first certified defense that has \emph{deterministic} robustness guarantees against adversarial point clouds. {\name} can transform an arbitrary point cloud classifier $\baseclf$ (called \emph{base point cloud classifier}) to be certifiably robust  against adversarial point clouds. Specifically, given a point cloud and a base point cloud classifier $\baseclf$, {\name} first divides the point cloud into multiple disjoint sub-point clouds using a hash function, then uses $\baseclf$ to predict a label for each sub-point cloud, and finally takes a majority vote among the predicted labels as the predicted label for the original point cloud. We prove that {\name} certifiably predicts the same label for a point cloud when the number of arbitrarily added, deleted, and/or modified points  is no larger than a threshold, which is known as \emph{certified perturbation size}. Moreover, we also prove that our derived certified perturbation size is \emph{tight}, i.e., without making assumptions on the base point cloud classifier $\baseclf$,  it is theoretically impossible to derive a  certified perturbation size for {\name} that is larger than ours. 

We consider three scenarios about how {\name} could be applied in practice and propose methods to optimize the performance of {\name} in these scenarios. In particular, we consider two parties: \emph{model provider} and \emph{customer}.  A model provider (e.g., Google, Meta) has enough labeled data and computation resource to train a base point cloud classifier $\baseclf$  and shares it with customers (e.g., a less resourceful company). Given $\baseclf$, a customer uses {\name} to classify its (adversarial) point clouds. We note that the model provider and customer can be the same entity, e.g., a company trains and uses $\baseclf$ itself. We consider three scenarios, in which  $\baseclf$ is trained by the model provider differently and/or used by a customer differently. 

Scenario I represents a naive application of {\name}, in which the base point cloud classifier $\baseclf$ is trained using a standard training algorithm and a customer directly applies {\name} to classify its point clouds based on $\baseclf$. {\name} achieves suboptimal performance in Scenario I because $\baseclf$, trained on point clouds, is not accurate at classifying sub-point clouds as they have different distributions.  Therefore, in Scenario II, we consider a model provider trains  $\baseclf$ to optimize the performance of {\name}. In particular, the model provider divides each training point cloud into multiple sub-point clouds following {\name} and trains $\baseclf$ based on sub-point clouds. In Scenario III, we consider the model provider has trained $\baseclf$ using a standard training algorithm (like Scenario I). However, instead of directly applying $\baseclf$ to classify sub-point clouds, a customer prepends a \emph{Point Completion Network (PCN)}~\cite{yuan2018pcn} to $\baseclf$. Specifically, a PCN takes a sub-point cloud as input and outputs a completed point cloud, which is then classified by $\baseclf$. Moreover, we propose a new loss function to train the PCN such that its completed point clouds are classified by  $\baseclf$ with higher accuracy, which further improves the performance of {\name}. 

We perform systematic evaluation on  ModelNet40 dataset~\cite{modelnet40} and two variants of ScanObjectNN dataset~\cite{scanobjectnn}. Our experimental results show that  {\name}  significantly  outperforms the state-of-the-art certified defenses (randomized smoothing~\cite{cohen2019certified} and PointGuard~\cite{liu2021pointguard}) even though their robustness guarantees are probabilistic. For instance, on ModelNet40 dataset, {\name} achieves a \emph{certified accuracy} of 79\% when an attacker can arbitrarily perturb at most 50 points in a point cloud, where certified accuracy is a lower bound of testing accuracy. Under the same setting, the certified accuracy of both randomized smoothing and PointGuard is $0$. We also extensively evaluate {\name} in the three application scenarios. 

In summary, we make the following contributions: (1) We propose {\name}, the first certified defense with deterministic robustness guarantees against adversarial point clouds. (2) We design multiple methods to optimize the performance of {\name} in multiple application scenarios. (3) We extensively evaluate {\name} and compare it with state-of-the-art certified defenses.

\section{Related Work}
\label{related_work}

Many works~\cite{xiang2019generating, zhao2020isometry, kim2021minimal,liu2020adversarial,wicker2019robustness, zheng2019pointcloud,yang2019adversarial,hamdi2020advpc,ma2020efficient,liu2022imperceptible} developed attacks to point cloud classification. Next, we discuss empirical and certified defenses against these attacks.

\noindent{\bf Empirical defenses:} Many empirical defenses~\cite{zhou2019dup,dong2020self,liu2019extending,yang2019adversarial,wu2020if,zhang2022pointcutmix,li2022robust,sun2022benchmarking} have been proposed to defend against adversarial point clouds. However, those empirical defenses do not have formal robustness guarantees and thus can often be broken by advanced, adaptive attacks. For instance, Sun et al.~\cite{sun2020adversarial} designed adaptive attacks with 100\% attack success rate to adversarial training based defenses~\cite{zhou2019dup,dong2020self}.

\noindent{\bf Certified defenses:} Randomized smoothing~\cite{cao2017mitigating,liu2018towards,lecuyer2019certified,li2018second,cohen2019certified,salman2019provably} can turn an arbitrary classifier into a certifiably robust one via adding random noise to an input. When generalized to point cloud, randomized smoothing can only certify robustness against point modification attacks~\cite{liu2021pointguard}.  PointGuard~\cite{liu2021pointguard} creates multiple sub-point clouds from a point cloud and takes a majority vote among them to predict the label of the point cloud. However, unlike {\name}, each sub-point cloud is sampled from the point cloud uniformly at random. Due to the inherent randomness, both randomized smoothing and PointGuard only have probabilistic guarantees. \cite{lorenz2021robustness,perez20223deformrs} proposed 3DCertify and 3DeformRS to certify robustness of point cloud classification against common 3D transformations, e.g., rotations. However, both methods are not applicable to point addition (or deletion or modification or perturbation) attacks, which can \emph{arbitrarily} manipulate points. Fischer et al.~\cite{fischer2021scalable} generalized randomized smoothing~\cite{cohen2019certified} to certify robustness of point cloud segmentation, which is different from our work since we focus on point cloud classification.

\section{Problem Definition}
In point cloud classification, a point cloud classifier $\pcc$ predicts a point cloud $\pointcloud$ into one of  $c$ classes (denoted as $\{1,2,\cdots,c\}$). Formally, we have $\pcc: \pointcloud \longrightarrow \{1,2,\cdots,c\}$. A point cloud $\pointcloud$ is a \emph{set} of points. For simplicity, we denote $\pointcloud = \{\mathbf{e}_1, \mathbf{e}_2,\cdots, \mathbf{e}_n\}$, where $n$ is the number of points. Each point $\mathbf{e}_i=(e_{i1},e_{i2},\cdots,e_{io})$ is a vector that specifies the three coordinates of the point in the three-dimensional space and (optionally) the point's other information  such as RGB values that describe the color features. 

\subsection{Adversarial Point Clouds}
Existing attacks to point cloud classification can be categorized into \emph{point addition attacks}~\cite{xiang2019generating,kim2021minimal,yang2019adversarial}, \emph{point deletion attacks}~\cite{wicker2019robustness,zheng2019pointcloud,yang2019adversarial}, \emph{point modification attacks}~\cite{xiang2019generating,kim2021minimal,yang2019adversarial,hamdi2020advpc}, and \emph{point perturbation attacks}~\cite{zhao2020isometry,yang2019adversarial,liu2020adversarial}. Specifically, in point addition (or deletion or modification) attacks, an attacker can \emph{arbitrarily} add new points (or delete or modify existing points) to a point cloud. Note that modifying a point is equivalent to deleting an existing point and adding a new point. In point perturbation attacks, an attacker can use any combination of the three operations (i.e., addition, deletion, and modification) to perturb a point cloud.  

Given a point cloud $\pointcloud$, we use $\pointcloud'$ to denote its adversarially perturbed version. We use $d(\pointcloud,\pointcloud')$ to denote the \emph{perturbation size}, i.e.,  the minimum number of perturbed (i.e., added, deleted, and/or modified) points that can turn $\pointcloud$ to $\pointcloud'$. Formally, we have $d(\pointcloud,\pointcloud')=$ $\max(|\pointcloud|,|\pointcloud'|)-|\pointcloud\cap\pointcloud'|$, where $|\cdot|$ measures the number of points in a point cloud and $\cap$ represents the intersection between two sets. Suppose we are given a perturbation size $t$. We use $\mathcal{S}(\pointcloud,t)$ to denote the set of all possible adversarial point clouds whose perturbation sizes are at most $t$. Formally, we have $\mathcal{S}(\pointcloud,t) = \{\pointcloud'|d(\pointcloud,\pointcloud')\leq t\}$.

\subsection{Certifiably Robust Point Cloud Classifier}
\vspace{-2mm}
\myparatight{Certified perturbation size} We say a point cloud classifier is certifiably robust if it certifiably predicts the same label for a point cloud when the number of points arbitrarily added, deleted, and/or modified by an attacker is less than a threshold, called \emph{certified perturbation size}. Formally, given a point cloud $\pointcloud$ and a point cloud classifier $\pcc$, we say $\pcc$ is certifiably robust for $\pointcloud$ with a certified perturbation size $t(\pointcloud)$ if $\pcc$ predicts the same label for the point cloud $\pointcloud$ and any adversarial point cloud with perturbation size at most $t(\pointcloud)$, i.e., $\pcc(\pointcloud')=\pcc(\pointcloud)$ for $\forall \pointcloud' \in \mathcal{S}(\pointcloud,t(\pointcloud))$. 

\myparatight{Probabilistic vs. deterministic guarantees} We say a point cloud classifier $\pcc$ produces an \emph{incorrect} certified perturbation size $t(\pointcloud)$ for a point cloud  $\pointcloud$ if there exists an adversarial point cloud $\pointcloud'$ with perturbation size at most $t(\pointcloud)$ such that $\pcc$ predicts different labels for $\pointcloud'$ and $\pointcloud$, i.e., $\exists \pointcloud' \in \mathcal{S}(\pointcloud,t(\pointcloud))$, $\pcc(\pointcloud')\neq \pcc(\pointcloud)$.  
A certifiably robust point cloud classifier has \emph{probabilistic} guarantees if it produces an incorrect certified perturbation size for a point cloud with an error probability $\alpha$. A certifiably robust point cloud classifier has \emph{deterministic} guarantees if its produced certified perturbation sizes are always correct.

\section{Our {\name}}
We first describe our  {\name} framework, which builds an ensemble point cloud classifier from an arbitrary point cloud classifier (called \emph{base point cloud classifier}). Then, we derive the certified perturbation size of our ensemble point cloud classifier. 

\subsection{Building an Ensemble Point Cloud Classifier}
\label{building_ensemble_pcc}
\vspace{-1mm}

\myparatight{Dividing a point cloud into multiple disjoint sub-point clouds} Suppose we have a point cloud $\pointcloud=\{\mathbf{e}_{1}, \mathbf{e}_{2}, \cdots, \mathbf{e}_{n}\}$, where $n$ is the number of points and $\mathbf{e}_i=(e_{i1},e_{i2},\cdots, e_{io})$ ($i=1,2,\cdots, n$) is a point. Our idea is to divide the point cloud $\pointcloud$ into $m$ sub-point clouds. In particular, our division aims to achieve three goals. The first goal is that an adversarially perturbed point should influence a small number of sub-point clouds. In other words, most sub-point clouds are not influenced when the number of adversarially perturbed points is small. The second goal is that a point should be assigned into a sub-point cloud deterministically. 
As we will see in the next subsection, the first two goals enable us to derive a deterministic certified perturbation size of {\name} for a point cloud. The third goal is that the sub-point clouds should contain similar number of points. In particular, if some  sub-point clouds contain (much) less number of points, then the base point cloud classifier may be more likely to misclassify them. As a result, our ensemble point cloud classifier is less accurate. As we will see in our experiments, the third goal enables {\name} to produce larger certified perturbation sizes. 

To reach the first goal, we propose to assign each point into one sub-point cloud. Therefore, an adversarially added or deleted point only influences one sub-point cloud, i.e., adding one point only influences the sub-point cloud  which the added point is assigned to while deleting one point only influences the sub-point cloud from which the point is deleted. An adversarially modified point influences at most two sub-point clouds, i.e., the sub-point clouds which the point belongs to before and after modification.  To reach the second goal, we propose to use the coordinates $e_{i1},e_{i2},\cdots,e_{io}$ to determine which sub-point cloud that the point $\mathbf{e}_i$ belongs to. Note that we cannot use the index of a point since the point cloud contains a \emph{set} of points. To reach the third goal, we propose to use a hash function to assign a point $\mathbf{e}_i$ into a sub-point cloud. While {\name} is applicable with any hash function,  we use a \emph{cryptographic hash function} (e.g., MD5) in our experiments because it is designed to have uniformly random output.  In particular, a cryptographic hash function takes any string as input and outputs a large integer that is roughly uniformly at random in the output space of the cryptographic hash function. 

Combining the above three ideas, we first transform each value $e_{ij}$ ($j=1,2,\cdots, o$) into a string $s_{ij}$, then concatenate $s_{ij}$'s of a point $\mathbf{e}_i$ into $S_i$ (i.e., $S_i=s_{i1} \oplus s_{i2}\oplus \cdots \oplus s_o$, where $\oplus$ represents string concatenation), and finally use a  hash function (denoted as $Hash$) to compute the hash value of $S_i$ (denoted as $Hash(S_i)$). We assign the point $\mathbf{e}_{i}$ to the $r_i$th sub-point cloud, where $r_i=Hash(S_i) \text{ mod } m$, where mod is the modulo operation. 
For simplicity, we use $\pointcloud_1,\pointcloud_2,\cdots,\pointcloud_m$ to denote the $m$ sub-point clouds created from $\pointcloud$. Note that some sub-point clouds may be empty, i.e., include no points. 

\myparatight{Building an ensemble point cloud classifier} Given the $m$ sub-point clouds $\pointcloud_1,\pointcloud_2,\cdots,\pointcloud_m$ created from the point cloud $\pointcloud$ and a base point cloud classifier $\baseclf$, we build an ensemble point cloud classifier $\ensclf$. In particular, we first use $\baseclf$ to predict a label for each non-empty sub-point cloud. Note that we do not consider those empty sub-point clouds. Then, we compute the number (denoted as $M_l(\pointcloud)$) of non-empty sub-point clouds that are predicted to have label $l$ by $\baseclf$. 
Formally, we define $M_l(\pointcloud) = \sum_{i=1}^{m}\mathbb{I}(\baseclf(\pointcloud_i)=l)\cdot \mathbb{I}(|\pointcloud_i|>0)$, where $l=1,2,\cdots, c$, $\mathbb{I}$ is an indicator function, and $|\cdot|$ measures the number of points in a sub-point cloud. For simplicity, we call $M_l(\pointcloud)$ \emph{label frequency} for label $l$. Our ensemble point cloud classifier $\ensclf$ predicts the label whose label frequency is the largest for the point cloud $\pointcloud$. Formally, we denote $\ensclf(\pointcloud)$ as the label predicted for $\pointcloud$ by $\ensclf$ and we have $\ensclf(\pointcloud) = \argmax_{l=1,2,\cdots,c} M_l(\pointcloud)$. 

We note that there may exist multiple labels with tied largest label frequencies. Usually, we break such ties uniformly at random, i.e., we predict a label among the tied ones uniformly at random. However, such random tie breaking introduces randomness and makes it hard to derive deterministic guarantees. To address the challenge, we break ties using the label indices deterministically. In particular, we order the $c$ labels as $1,2,\cdots,c$ and we predict the ``smallest'' label among the tied ones. 
 For example, suppose labels 1 and 2 have tied largest label frequencies, i.e.,  $M_1(\pointcloud)=M_2(\pointcloud)>M_l(\pointcloud)$, where $l\neq 1,2$. Our $\ensclf$ predicts label $1$ for $\pointcloud$. More formally, our $\ensclf$ predicts label $y$ for a point cloud $\pointcloud$ if $M_{y}(\pointcloud) \geq \max_{l\neq y} (M_{l}(\pointcloud)+\mathbb{I}(y>l))$. 
 
\subsection{Deriving Certified Perturbation Size}
\vspace{-2mm}
\myparatight{Derivation goal} Suppose our ensemble point cloud classifier predicts a label $y$ for a point cloud $\pointcloud$. 
$\pointcloud'$ is an adversarially perturbed version of $\pointcloud$. Our goal is to derive the largest certified perturbation size $t(P)$ such that our ensemble point cloud classifier is guaranteed to predict label $y$ for any $\pointcloud'$ with perturbation size at most $t(P)$. Formally, we aim to find the largest $t(P)$ such that we have $M_{y}(\pointcloud') \geq \max_{l\neq y} (M_{l}(\pointcloud')+\mathbb{I}(y>l))$ for any $\pointcloud' \in \mathcal{S}(\pointcloud,t(\pointcloud))$. Our idea is to first derive a lower bound of $M_{y}(\pointcloud')$ and an upper bound of $\max_{l\neq y} (M_{l}(\pointcloud')+\mathbb{I}(y>l))$, and then find the largest $t(P)$ such that the lower bound is no smaller than the upper bound. Next, we first describe how we derive the lower/upper bounds and then how we find the largest certified perturbation size $t(P)$. 

\vspace{-1mm}
\myparatight{Deriving a lower bound of $M_{y}(\pointcloud')$ and an upper bound of $\max_{l\neq y} (M_{l}(\pointcloud')+\mathbb{I}(y>l))$} Recall that we divide a (adversarial) point cloud into $m$ sub-point clouds. Since each point only appears in one sub-point cloud, an adversarially added or deleted point only impacts one sub-point cloud and may change the label predicted by the base point cloud classifier for the impacted sub-point cloud. Moreover, a modified point only impacts two sub-point clouds at most and thus impacts the predicted labels for two sub-point clouds at most. For simplicity, we define an \emph{impact factor} $\tau$ for an operation (i.e., addition, deletion, modification) as the largest number of sub-point clouds that are impacted when the operation is applied to one point. The \emph{impact factor} is 1 for addition/deletion and 2 for modification. 

If an attacker can arbitrarily add (or delete or modify) at most $t$ points to $\pointcloud$, then at most $\tau \cdot t$ sub-point clouds in $\pointcloud_1, \pointcloud_2, \cdots, \pointcloud_m$ are impacted. Therefore, we have $M_y(\pointcloud')\geq M_y(\pointcloud)-\tau\cdot t$ and  $M_l(\pointcloud')\leq M_l(\pointcloud)+\tau\cdot t$ for $\forall l \neq y$ and $\forall \pointcloud' \in \mathcal{S}(\pointcloud,t)$. We treat $M_y(\pointcloud)-\tau\cdot t(\pointcloud)$ as a lower bound of $M_y(\pointcloud')$ and $\max_{l\neq y}  (M_l(\pointcloud)+\tau\cdot t + \mathbb{I}(y>l) )$ as an upper bound of $\max_{l\neq y} (M_{l}(\pointcloud')+\mathbb{I}(y>l))$.

\myparatight{Computing certified perturbation size}  Our goal is to find the largest $t$ such that the lower bound of $M_y(\pointcloud')$ is no smaller than the upper bound of $\max_{l\neq y} (M_{l}(\pointcloud')+\mathbb{I}(y>l))$. In other words, we aim to find the largest $t$ such that $M_y(\pointcloud)-\tau\cdot t \geq \max_{l\neq y} (M_l(\pointcloud)+\tau\cdot t+\mathbb{I}(y>l))$. Therefore, we have
$t \leq \frac{M_y(\pointcloud)-\max_{l\neq y} (M_l(\pointcloud)+\mathbb{I}(y>l))}{2\cdot \tau}$.
Since the number of points that an attacker can add (or delete or modify) should be an integer, we have the certified perturbation size as $t(\pointcloud) = \lfloor \frac{M_y(\pointcloud)-\max_{l\neq y} (M_l(\pointcloud)+\mathbb{I}(y>l))}{2\cdot \tau} \rfloor$, where $\lfloor \cdot \rfloor$ is the floor function. In summary, we have the following: 
\begin{theorem}[Certified Perturbation Size]
\label{main_theorem}
Suppose we have a point cloud $\pointcloud$, a hash function to divide $\pointcloud$ into $m$ disjoint sub-point clouds, a base point cloud classifier $\baseclf$, and label frequency $M_l(\pointcloud)$, where $l=1,2,\cdots,c$. Our ensemble point cloud classifier $\ensclf$ predicts the same label for $\pointcloud$ and its adversarially perturbed version $\pointcloud'$ once the perturbation size is at most $t(\pointcloud)$. Formally,  we have:
\begin{align}
\label{theorem1_theoretic_guarantee}
    \ensclf(\pointcloud') = \ensclf(\pointcloud)=y, \forall \pointcloud' \in \mathcal{S}(\pointcloud, t(\pointcloud)), 
\end{align}
where $t(\pointcloud) = \lfloor \frac{M_y(\pointcloud)-\max_{l\neq y} (M_l(\pointcloud)+\mathbb{I}(y>l))}{2\cdot\tau} \rfloor$. The impact factor $\tau$ is 1 for point addition and deletion attacks, while it is 2  for point modification and perturbation attacks since a point perturbation attack can use any combination of the three operations.
\end{theorem}

We also prove that our derived certified perturbation size is tight, i.e., without making any assumptions on the base point cloud classifier, it is theoretically impossible to derive a certified perturbation size that is larger than ours. Formally, we have the following theorem:
\begin{theorem} [Tightness]
\label{tightnesstheorem}
Given a point cloud $\pointcloud$ and a hash function to divide $\pointcloud$ into $m$ disjoint sub-point clouds, there exists an adversarial point cloud $\pointcloud' \in \mathcal{S}(\pointcloud, t(\pointcloud)+1)$ and a base point cloud classifier $\baseclf'$ such that our ensemble classifier predicts different labels for $\pointcloud$ and $\pointcloud'$. Formally, we have $\ensclf'(\pointcloud') \neq \ensclf'(\pointcloud)$, 
where $\ensclf'$ is the ensemble point cloud classifier built based on $\baseclf'$.
\end{theorem}
\vspace{-5mm}
\begin{proof}
See Appendix~\ref{proof_of_theorem2}.
\end{proof}

\section{Applications in Three Scenarios}

\vspace{-2mm}
\myparatight{Scenario I} This scenario is a naive application of {\name}. 
Suppose a model provider has trained a base point cloud classifier $\baseclf$ using the standard training algorithm, and shares it with customers in a black-box or white-box setting. In the black-box setting, the model provider only provides a prediction API for a customer, who can send a point cloud to the model provider and obtain its prediction made by $\baseclf$. In the white-box setting, the model provider shares the model parameters with a customer, who can use $\baseclf$ to classify point clouds locally.

Given a black-box or white-box access to $\baseclf$, a customer directly uses {\name} to classify its point clouds. Specifically, given a point cloud, the customer  first divides it into sub-point clouds, then uses $\baseclf$ to predict a label for each non-empty sub-point cloud, and finally takes a majority vote among the predicted labels of the sub-point clouds as the predicted label for the  point cloud. 

\vspace{-1mm}
\myparatight{Scenario II}
In Scenario I,  $\baseclf$ is trained on point clouds, and thus may be inaccurate to classify sub-point clouds as  they have different distributions with point clouds. As a result, {\name} is less accurate.  In  Scenario II, we consider that the model provider trains its $\baseclf$ to optimize the performance of {\name}. In particular, to make $\baseclf$ more accurate in classifying  sub-point clouds, we propose that the model provider  trains $\baseclf$ on sub-point clouds. In particular, the model provider divides each training point cloud into $m$ disjoint sub-point clouds following {\name} and uses the label of the training point cloud as the label of each sub-point cloud. Then, the model provider  trains $\baseclf$ on those labeled sub-point clouds. Similar to Scenario I, the model provider can share $\baseclf$ with a customer in a black-box or white-box setting, and a customer can directly use {\name} to classify point clouds based on $\baseclf$. 

\begin{figure}[!t]
    \centering
    \includegraphics[width =0.9\linewidth]{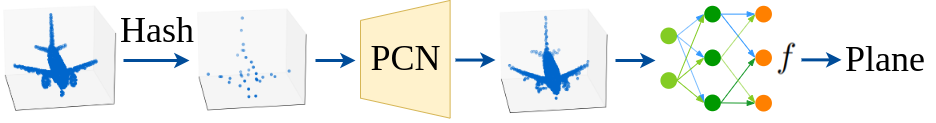}
    \caption{Composition of PCN and $\baseclf$ in Scenario III.}
    \label{fig:scenarioIII}
    \vspace{-4mm}
\end{figure}

\vspace{-1mm}
\myparatight{Scenario III} Similar to Scenario I,  we consider the model provider has trained $\baseclf$ using a standard training algorithm. However, instead of directly using $\baseclf$ to classify sub-point clouds, a customer 
adds a {\emph{Point Completion Network (PCN)}~\cite{yuan2018pcn}} before $\baseclf$ to improve its accuracy for sub-point clouds. In particular, the PCN takes a sub-point cloud as input and outputs a completed point cloud, which is then classified by $\baseclf$, as shown in Figure~\ref{fig:scenarioIII}.

\emph{\bf Formulating PCN learning as an optimization problem.} Suppose a customer has a set of unlabeled point clouds and (optionally) a small amount of labeled ones. The customer constructs a training dataset $\mathcal{D}_u$. Specifically, the customer divides each unlabeled point cloud into $m$ disjoint sub-point clouds following \name. $\mathcal{D}_u$ consists of a set of pairs $(\pointcloud_s,\pointcloud_p)$, where $\pointcloud_s$ is a sub-point cloud and $\pointcloud_p$ is the corresponding point cloud. A PCN takes $\pointcloud_s$ as input and aims to output $\pointcloud_p$. In existing point completion methods~\cite{yuan2018pcn, liu2020morphing, xie2020grnet, wen2020point, huang2020pf, wang2020cascaded, yu2021pointr}, learning a {PCN $\mathcal{C}$} essentially formulates a loss term $L_p(\mathcal{D}_u, \mathcal{C})$ over the training dataset 
$\mathcal{D}_u$ and then uses stochastic gradient descent (SGD) to minimize the loss. {We adopt the popular Chamfer Distance proposed in Fan et al. \cite{fan2017point} as the loss term in our experiments (the details can be found in Appendix).} 

However,  $\baseclf$ is still likely to misclassify the point clouds completed by such a PCN. The reason is that existing point completion methods did not aim to complete point clouds that can be classified by $\baseclf$ with high accuracy, since it is not their goal. To bridge this gap, we propose another loss term, which is smaller if the completed point clouds can be classified by $\baseclf$ with higher accuracy. Formally, we define the following loss term: $L_c(\mathcal{D}_l, \mathcal{C}, \baseclf) =\frac{1}{|\mathcal{D}_l|} \sum_{(\pointcloud,y)\in\mathcal{D}_l} \mathcal{L}(\baseclf(\mathcal{C}(\pointcloud)),y)$, where $\mathcal{D}_l$ is the set of labeled point clouds and $\mathcal{L}$ is the loss function for classification such as cross-entropy loss. Combining the two loss terms, our final loss  used to train a PCN is as follows:
\begin{align}
\label{final_optimization_problem}
    L_p(\mathcal{D}_u,\mathcal{C}) + \lambda \cdot L_c(\mathcal{D}_l, \mathcal{C},\baseclf),
\end{align}
where $\lambda$ is a hyperparameter used to balance the two loss terms. We note that Scenario III is not applicable to the customer without any unlabeled or labeled point clouds. Moreover, when a customer only has unlabeled point clouds, the customer can only use a standard point completion method to learn a PCN, i.e., $\lambda=0$ in Equation~\ref{final_optimization_problem}.

\begin{figure*}
    \centering
    \subfloat[Point addition attacks]{\includegraphics[width =0.23\textwidth]{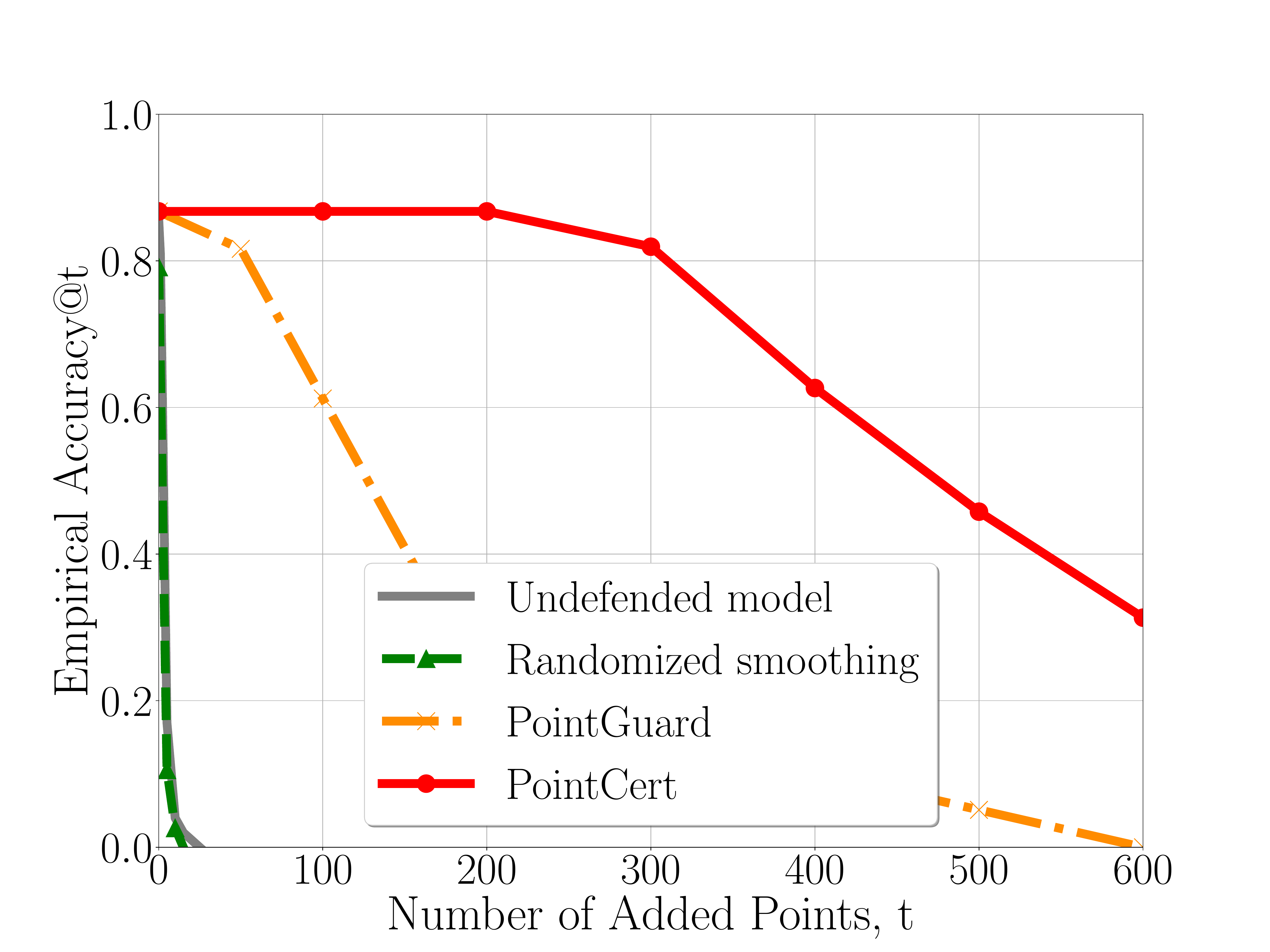}}    
    \subfloat[Point deletion attacks]{\includegraphics[width =0.23\textwidth]{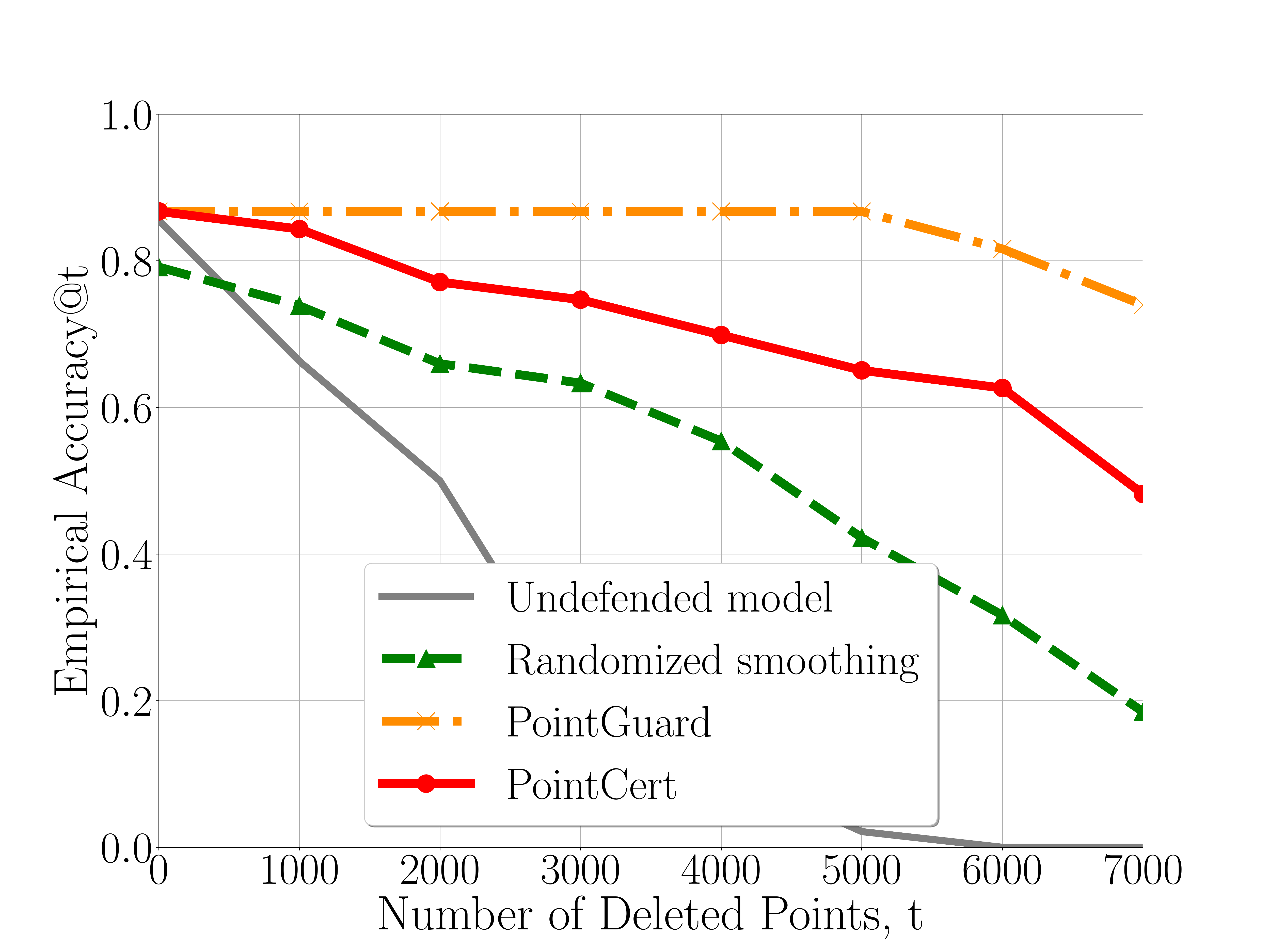}}
    \subfloat[Point modification attacks]{\includegraphics[width =0.23\textwidth]{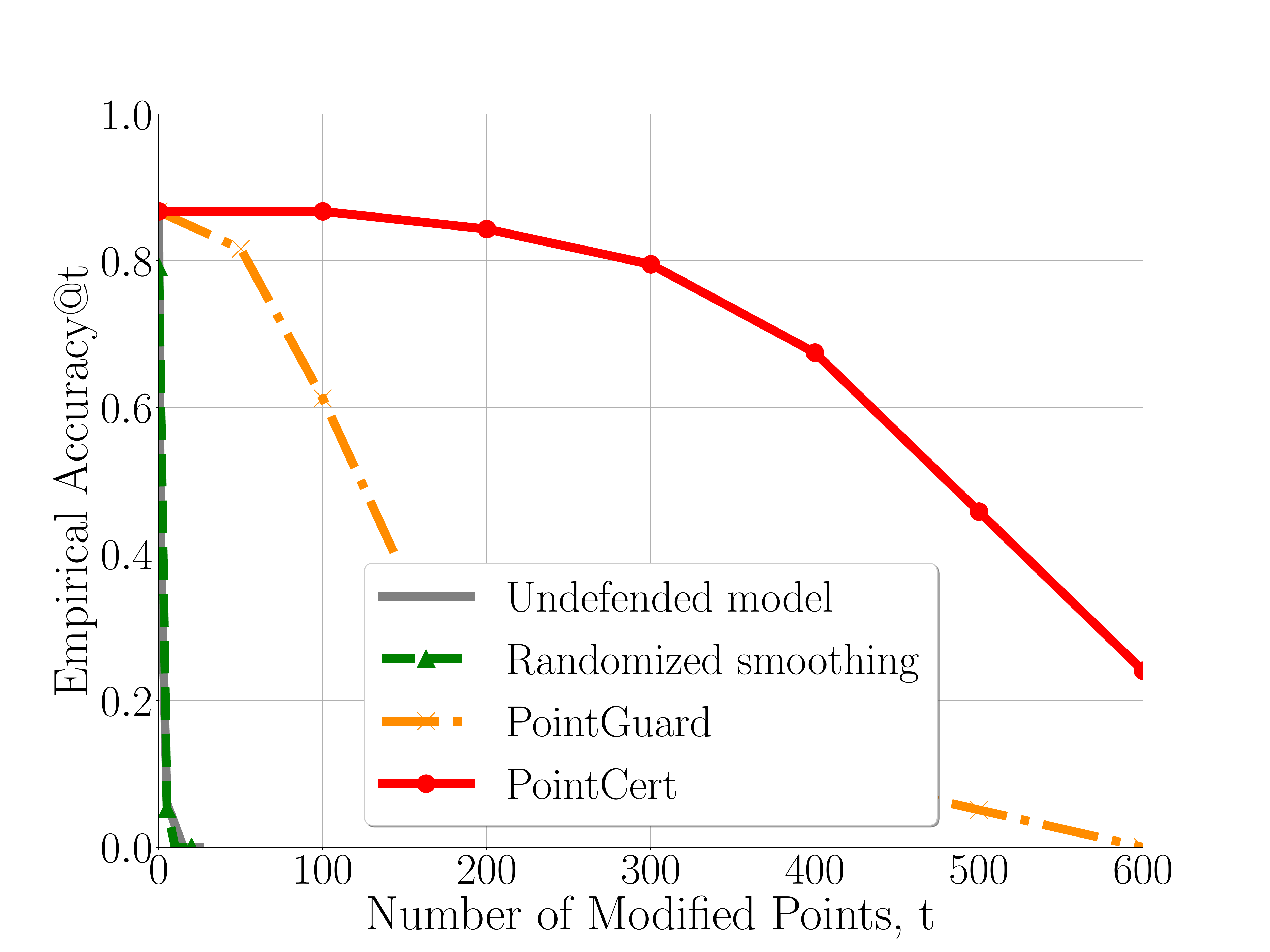}}
    \subfloat[Point perturbation attacks]{\includegraphics[width =0.23\textwidth]{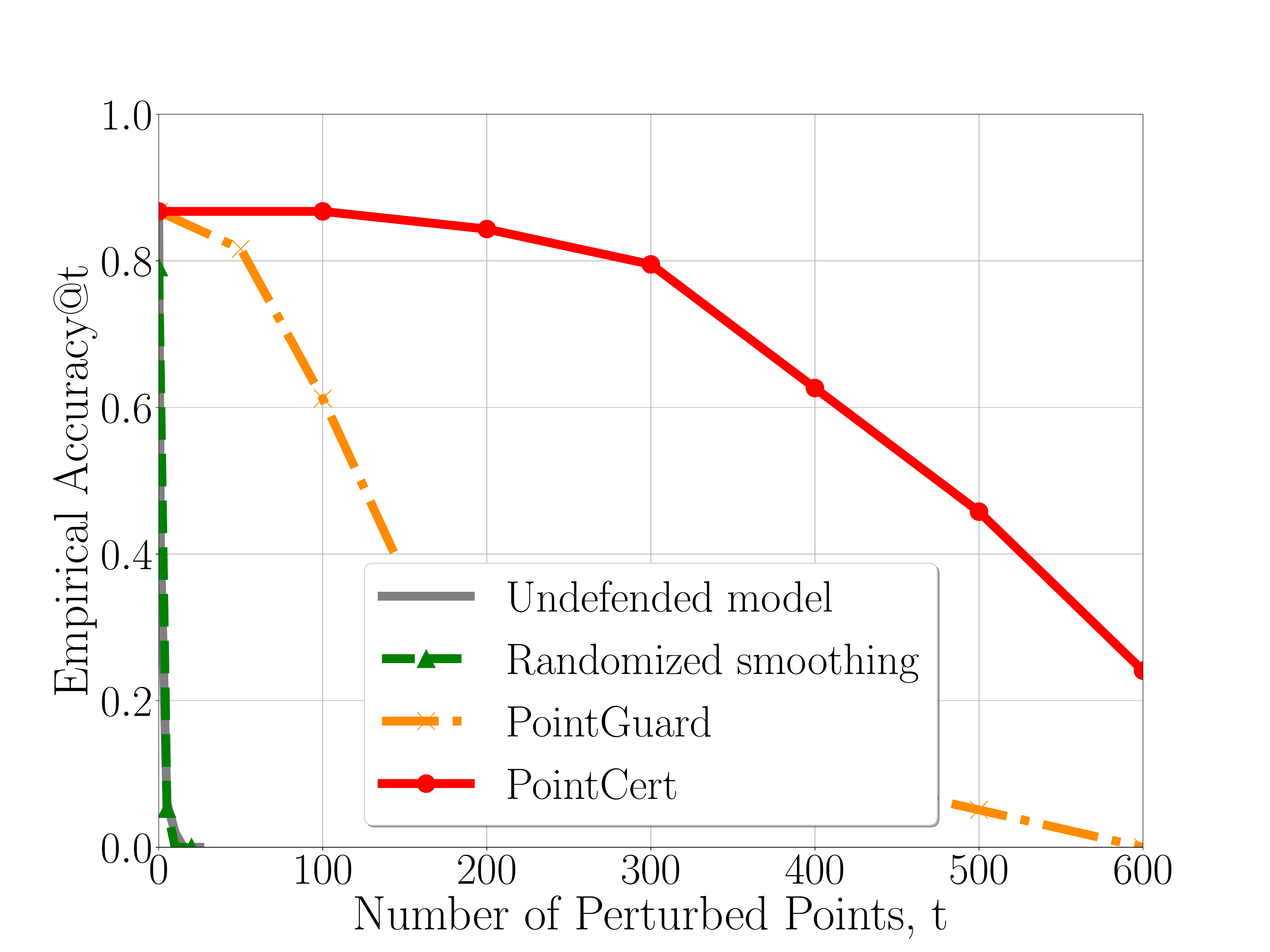}}
 \vspace{-2mm}
    \caption{Comparing the empirical accuracy of different defenses. Scenario II is considered.}
    \label{fig:compare_different_defenses_empirical_acc}
    \vspace{-2mm}
\end{figure*}
 
\begin{figure*}
    \centering
    \subfloat[Point addition attacks]{\includegraphics[width =0.23\textwidth]{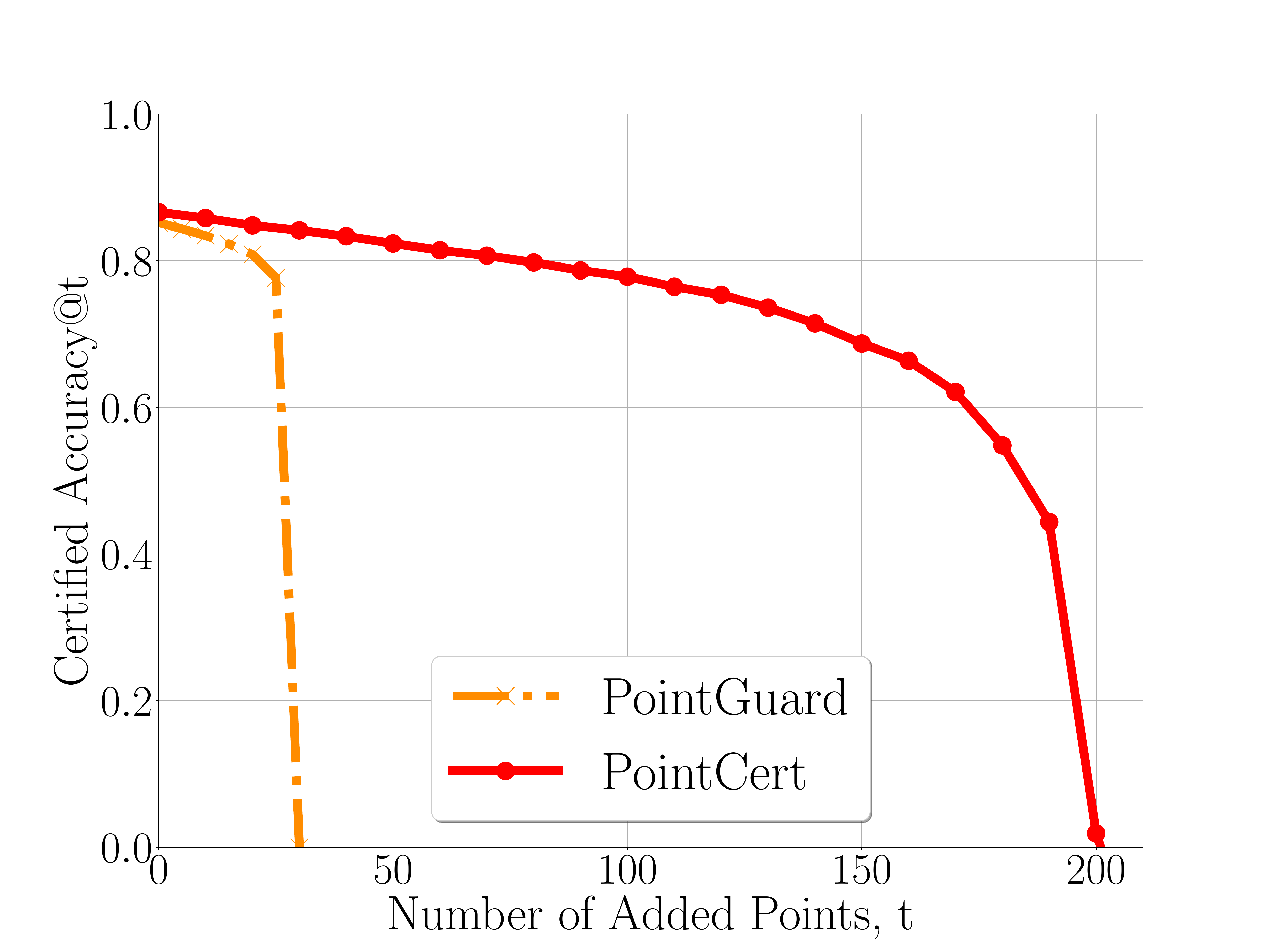}}    
    \subfloat[Point deletion attacks]{\includegraphics[width =0.23\textwidth]{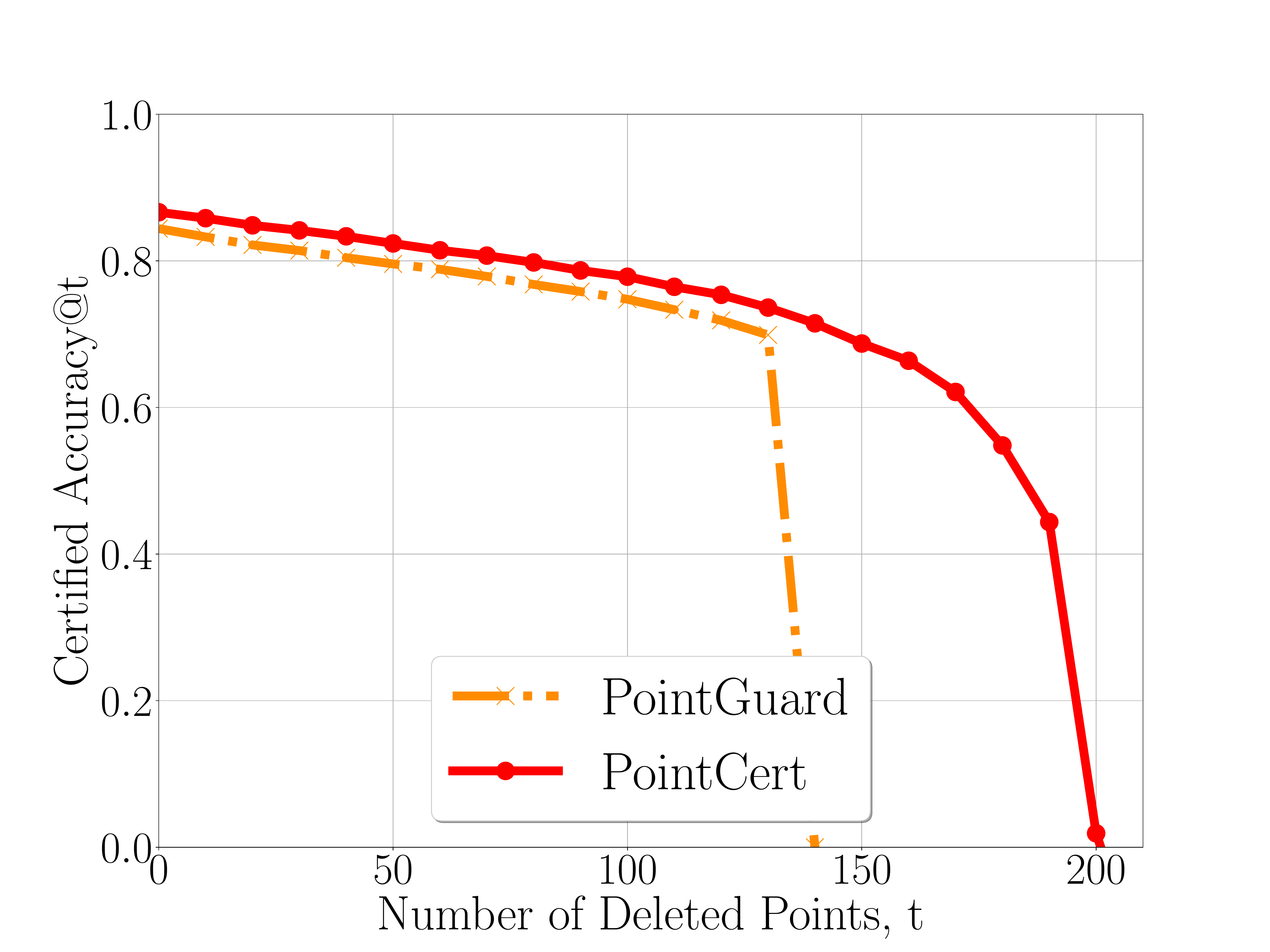}}
    \subfloat[Point modification attacks]{\includegraphics[width =0.23\textwidth]{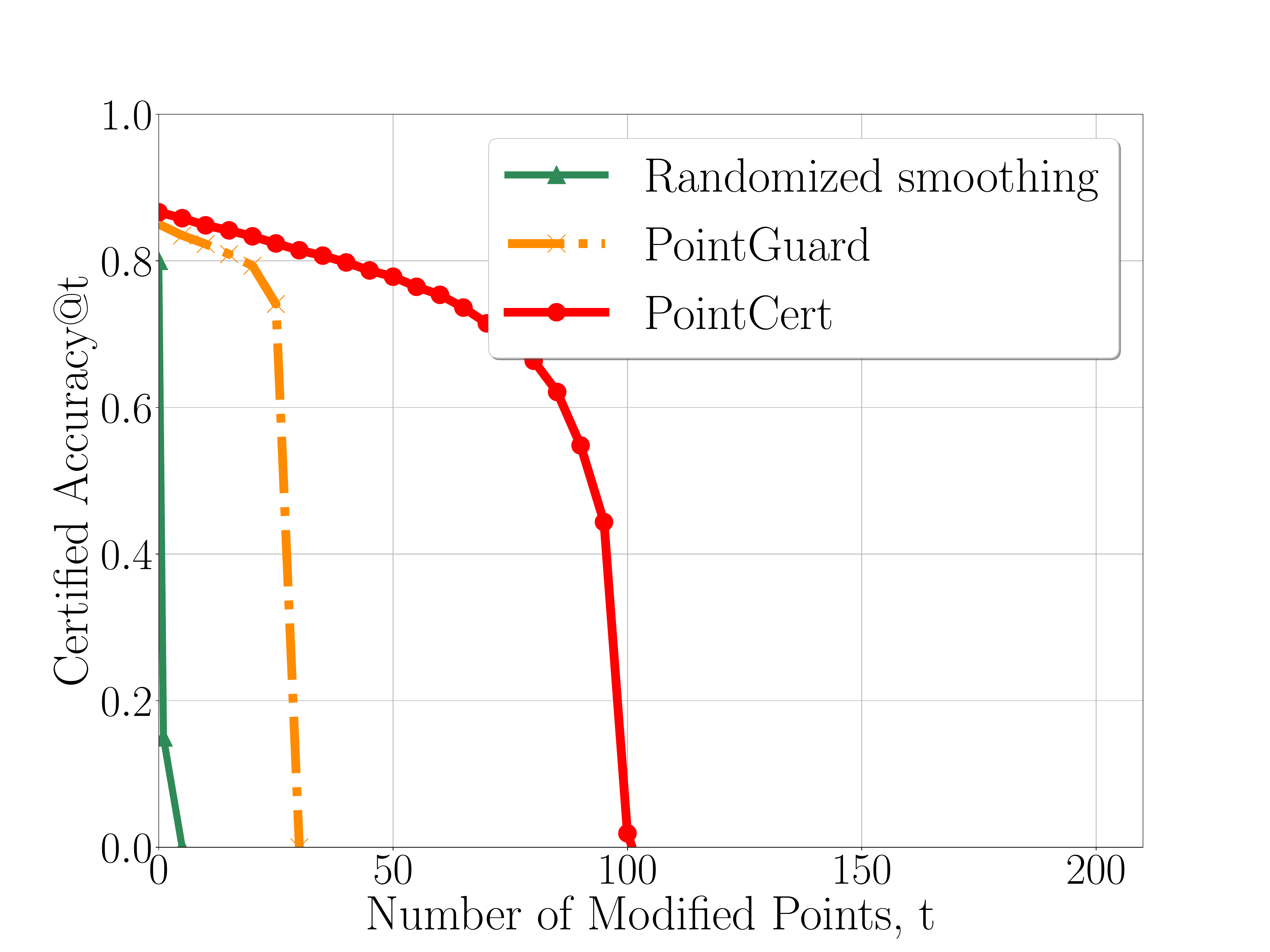}}
    \subfloat[Point perturbation attacks]{\includegraphics[width =0.23\textwidth]{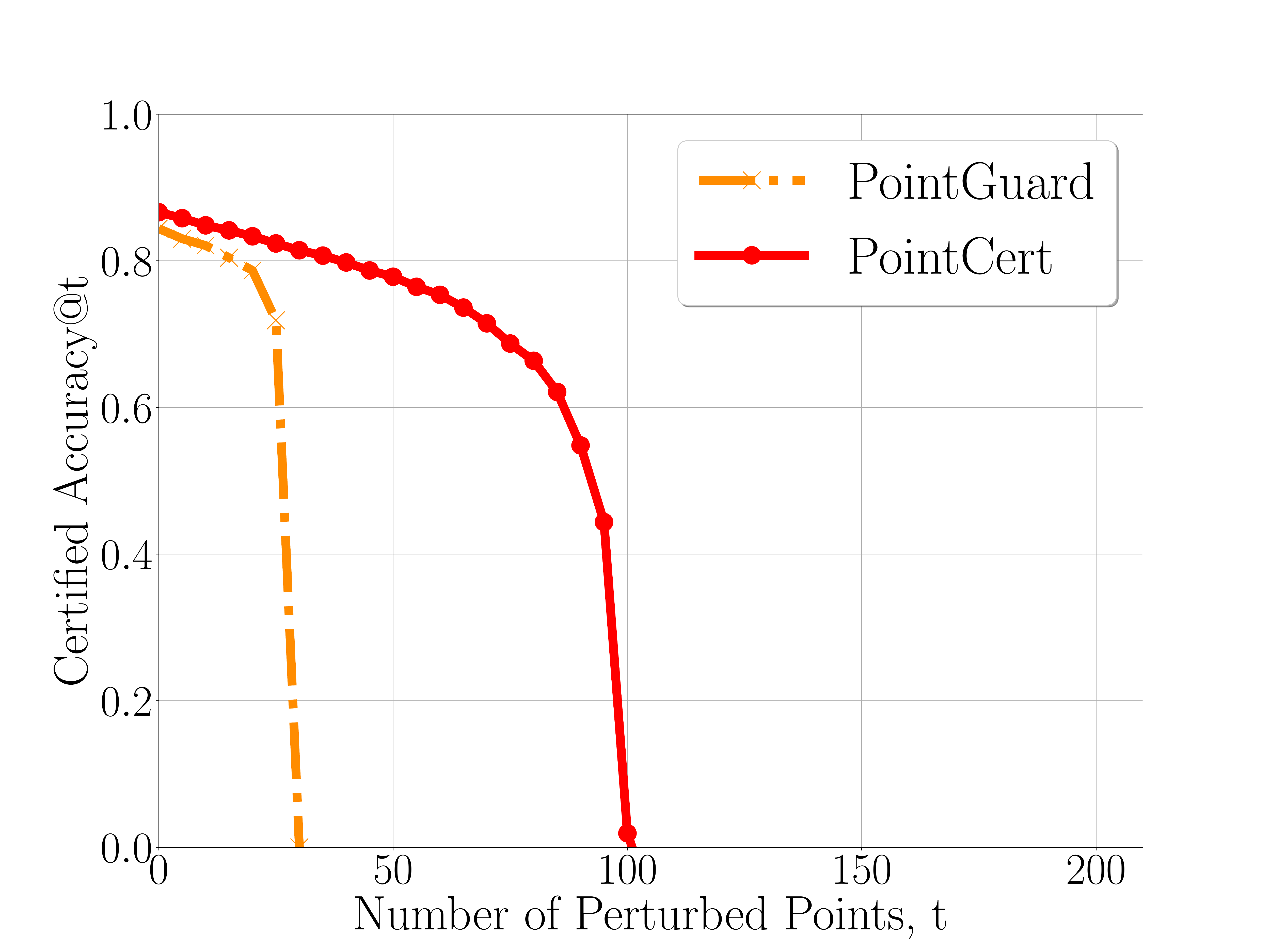}}
 \vspace{-2mm}
    \caption{Comparing the certified accuracy of randomized smoothing, PointGuard, and {\name}. Randomized smoothing can only provide certified robustness guarantees against point modification attacks. Scenario II is considered.}
    \label{fig:compare_different_defenses_certified_acc}
    \vspace{-3mm}
\end{figure*}

\emph{\bf Solving the optimization problem in white-box and black-box settings.} In the white-box setting, a customer has access to the model parameters of $\baseclf$. Therefore, the customer can solve the optimization problem in Equation~(\ref{final_optimization_problem}) using the standard SGD to learn a PCN. In the black-box setting, the customer only has access to the prediction API of $\baseclf$, and thus cannot solve the optimization problem using SGD. The customer could use zeroth-order optimization methods~\cite{ghadimi2013stochastic} to solve the optimization problem. However, such method often incurs a large number of queries to the prediction API. To address the challenge, we propose that the customer learns a student model using \emph{knowledge distillation}~\cite{hinton2015distilling} by viewing $\baseclf$ as a teacher model. Roughly speaking, the customer can first query $\baseclf$ using his/her unlabeled and labeled point clouds (excluding labels) to obtain their {output logits  predicted by $\baseclf$, then divide the logits by $T$ which is a temperature parameter in knowledge distillation,} and finally train a student model. Given the student model, the customer can use it to replace  $\baseclf$ in Equation~(\ref{final_optimization_problem}) and train a PCN using SGD. We note that the customer essentially treats the composition of the PCN and the student model (or the teacher model) as a new base point cloud classifier in our {\name} framework.

\section{Experiments}
\subsection{Experimental Setup}
\vspace{-2mm}

\myparatight{Datasets and models} We adopt two publicly available benchmark datasets, namely ModelNet40~\cite{wu20153d} and two variants of ScanObjectNN~\cite{uy2019revisiting}, in our evaluation. Each point cloud of ModelNet40 has 10,000 points and we also keep at most 10,000 points in each point cloud of ScanObjectNN. We do not reduce the size of a point cloud by sub-sampling its points to simulate real-world attack scenarios. Our method and compared baselines are evaluated using same number of points. The detailed dataset description is shown in Appendix~\ref{app:dataset}. We evenly split the training point clouds in each dataset into two \textbf{balanced halves}. One half is used for the \textbf{model provider} to train base point cloud classifier in the three scenarios, and the other is used for a \textbf{customer} to train a PCN in Scenario III. We consider PointNet~\cite{qi2017pointnet} and DGCNN~\cite{wang2019dynamic}, which are frequently used by the community, as the base point cloud classifiers.

\myparatight{Compared methods} We compare {\name} with undefended model, randomized smoothing~\cite{cohen2019certified}, and PointGuard~\cite{liu2021pointguard}. Both randomized smoothing and PointGuard only have probabilistic robustness guarantees. Details of these methods can be found in Appendix~\ref{app:method}.

\myparatight{Evaluation metrics} 
We use \emph{Empirical Accuracy@$t$} and \emph{Certified Accuracy@$t$}  as evaluation metrics. In particular, Certified Accuracy@$t$ is the fraction of testing point clouds in a testing dataset whose certified perturbation sizes are at least $t$ and whose labels are correctly predicted. The Empirical Accuracy@$t$ is the testing accuracy of each model under the empirical attacks with perturbation size $t$. We note that the Certified Accuracy@$t$ is a \emph{lower bound} of testing accuracy that a defense can achieve when the perturbation size is at most $t$, no matter how the perturbation is crafted.
Empirical Accuracy@$t$ is an \emph{upper bound} of testing accuracy that each model can achieve under attacks with perturbation size at most $t$.
For undefended model, we only report Empirical Accuracy@$t$ because it does not have certified robustness guarantees. Besides, randomized smoothing can only provide certified robustness guarantees for point modification attacks, though we can report its Empirical Accuracy@$t$ against other attacks.

In experiments, we use the attacks developed by~\cite{xiang2019generating} for point addition, modification, and perturbation attacks and ~\cite{wicker2019robustness} for point deletion attack. We note that there are no existing adversarial point cloud attacks tailored to randomized smoothing, PointGuard, and {\name}. To bridge this gap, we generalize existing attacks to these ensemble models and compute their Empirical Accuracy@$t$. The key idea of our attacks to ensemble models is to identify a set of critical points to add (or delete) such that the classification losses of the base point cloud classifier on different groups of point clouds (e.g., sub-point clouds in {\name}) are maximized. Details of our empirical attacks can be found in Appendix~\ref{app:attack}.

\myparatight{Parameter setting} Our {\name} has a parameter $m$ and a hash function to divide a point cloud into $m$ sub-point clouds. By default, we set $m=400$ and use MD5 as the hash function. Despite the large $m$, the inference time per testing point cloud of {\name} is less than 0.51s  because the sub-point cloud sizes are small. Moreover, we set PointNet as the default base point cloud classifier. In Scenario III, we set the default value of $\lambda$ to be $5 \times 10^{-4}$ to balance the two loss terms. By default, we assume 25\% of the customer's point clouds are labeled while the remaining is unlabeled in Scenario III. For point cloud completion, we use coarse output of PCN \cite{yuan2018pcn} since we do not require the  fine-grained output. In Scenario I and II, white-box and black-box settings have no difference. In Scenario III, we assume the white-box setting by default; and in the black-box setting, we learn a student model using knowledge distillation with a temperature $T=20$. Due to space constraint, we show the results on ScanObjectNN in Appendix. 

\subsection{Experimental Results}
\subsubsection{Comparing Different Defenses}

Figure~\ref{fig:compare_different_defenses_empirical_acc} compares the empirical accuracy of all methods under empirical attacks, while Figure~\ref{fig:compare_different_defenses_certified_acc} compares the certified accuracy of randomized smoothing, PointGuard, and {\name} in Scenario II. We note that these defenses have accuracy-robustness tradeoffs, which are controlled by their parameters, e.g., $m$ in {\name}. Therefore, to fairly compare {\name} with randomized smoothing and PointGuard, we make them have similar certified accuracy under no attacks (i.e., $t=0$) by following previous work~\cite{liu2021pointguard}. In particular,  we use the default $m$ for {\name} and respectively search $\sigma$ and $k$ for randomized smoothing and PointGuard. Our searched parameters are $\sigma =0.25$ and $k=256$. We use the settings in~\cite{liu2021pointguard} for other parameters of randomized smoothing and PointGuard. 

\begin{figure}
    \centering
    \subfloat[Point addition/deletion\\ attacks]{\includegraphics[width =0.23\textwidth]{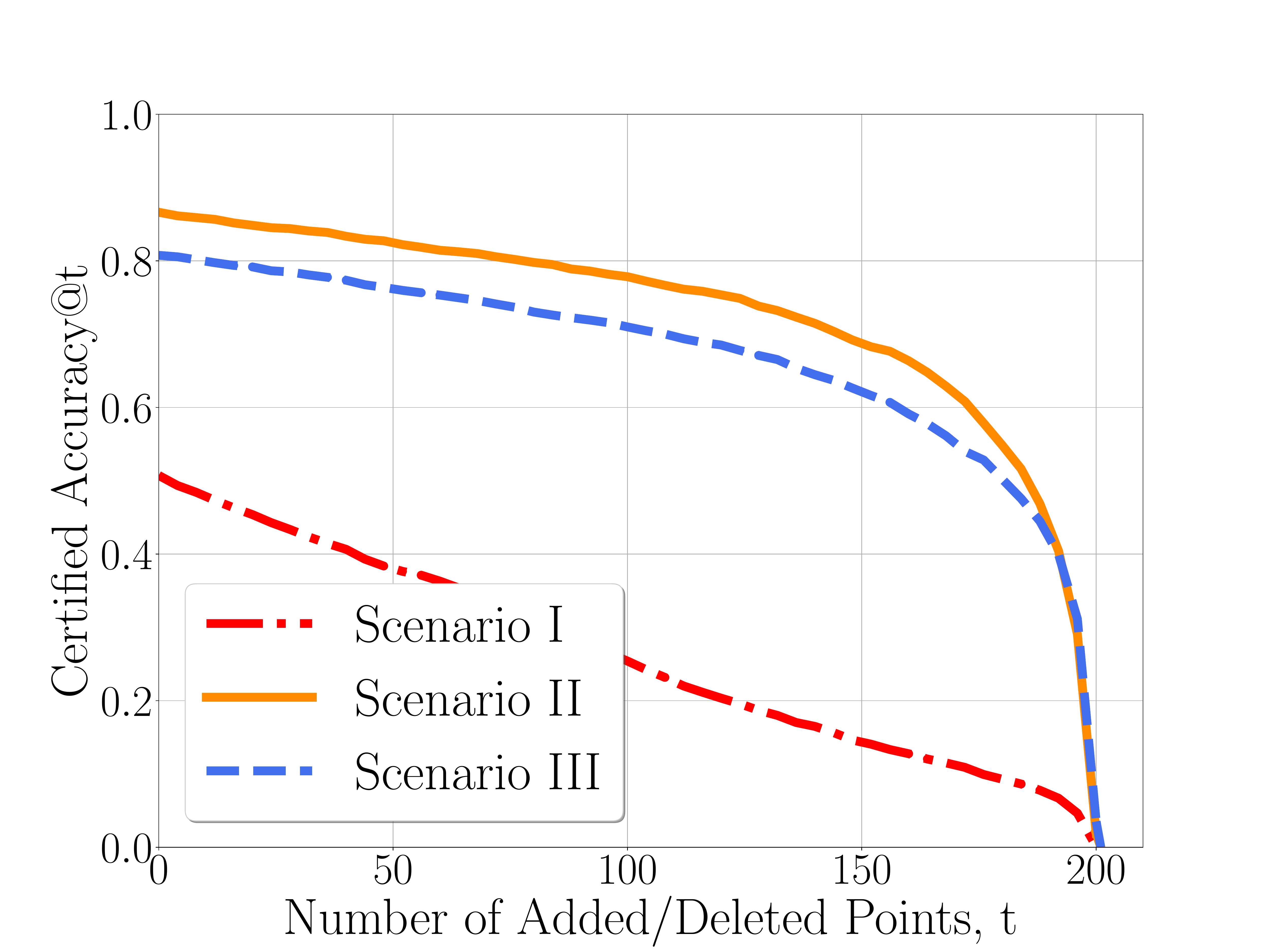}\label{fig:threescenarios-modelnet40-a}}    
    \subfloat[Point modification/perturbation attacks]{\includegraphics[width =0.23\textwidth]{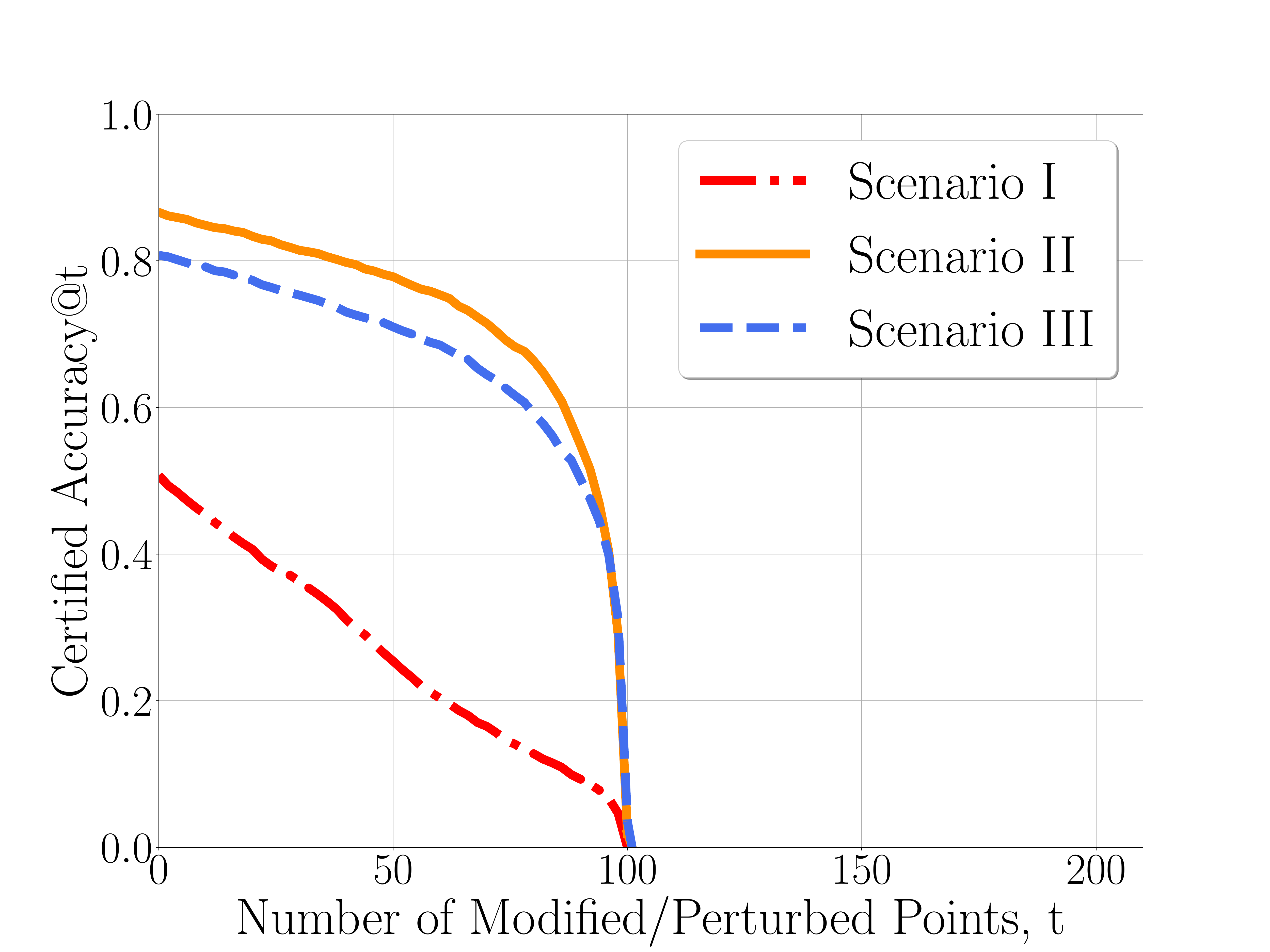}\label{fig:threescenarios-modelnet40-b}}
\vspace{-2mm}
    \caption{Comparing the certified accuracy of {\name} in the three application scenarios under different attacks. }
    \label{fig:threescenarios-modelnet40}
    \vspace{-5mm}
\end{figure}

We have the following observations from the experimental results. First, an undefended model is not robust against adversarial point clouds. For instance,  adding or modifying only 1 out of 10,000 points can substantially reduce its empirical accuracy. Second, {\name} achieves larger empirical accuracy and certified accuracy than randomized smoothing and PointGuard even though their certified robustness guarantees are probabilistic. The reasons are that 1) randomized smoothing adds Gaussian noise to every point in a point cloud, making its classification less accurate, and 2) an adversarially added, deleted, and/or modified point impacts multiple subsampled point clouds in PointGuard. In contrast,  {\name}  does not add noise to points in a point cloud and each perturbed point impacts at most 1 or 2 sub-point clouds. 

An interesting exception is that PointGuard achieves better empirical accuracy than {\name} under our empirical point deletion attacks. The reason is that each sub-point cloud in {\name} contains much less number of points after thousands of points are deleted, and thus the base point cloud classifier in {\name} is less accurate. In contrast, each subsampled point cloud in PointGuard still contains $k$ points even if thousands of points are deleted. Third, every method achieves much higher empirical accuracy against point deletion attacks than against other attacks when the perturbation size is the same, which indicates that state-of-the-art point deletion attack is not powerful enough.

\vspace{-2mm}
\subsubsection{Comparing the Three Scenarios}
\vspace{-2mm}
Figure~\ref{fig:threescenarios-modelnet40} compares the three application scenarios of {\name} under attacks.
In each scenario, the certified accuracy of {\name} is the same for point addition and deletion attacks, and is the same for point modification and perturbation attacks. 
Thus, both Figures~\ref{fig:threescenarios-modelnet40-a} and~\ref{fig:threescenarios-modelnet40-b} showcase the certified accuracy of {\name} under two attacks.

First, {\name} achieves the best certified accuracy in Scenario II as the base point cloud classifier in Scenario II is trained on sub-point clouds and is more accurate in classifying them. Second, {\name} achieves better certified accuracy in Scenario III than in Scenario I. The reason is that, in Scenario III, a customer trains a PCN to turn a sub-point cloud into a completed point cloud, which can be well classified by the base point cloud classifier trained using a standard algorithm. 
Third, in each scenario, given the same certified accuracy, the perturbation size that {\name} can tolerate under point addition/deletion attacks is twice of that under point modification/perturbation attacks. 
The reason is that modifying a point is equivalent to adding a point and deleting a point, which could impact two sub-point clouds in the worst case. Due to such relationship, we compare results on point addition attacks in the following section.

\begin{figure}
    \centering
\subfloat[]{ \includegraphics[width =0.23\textwidth]{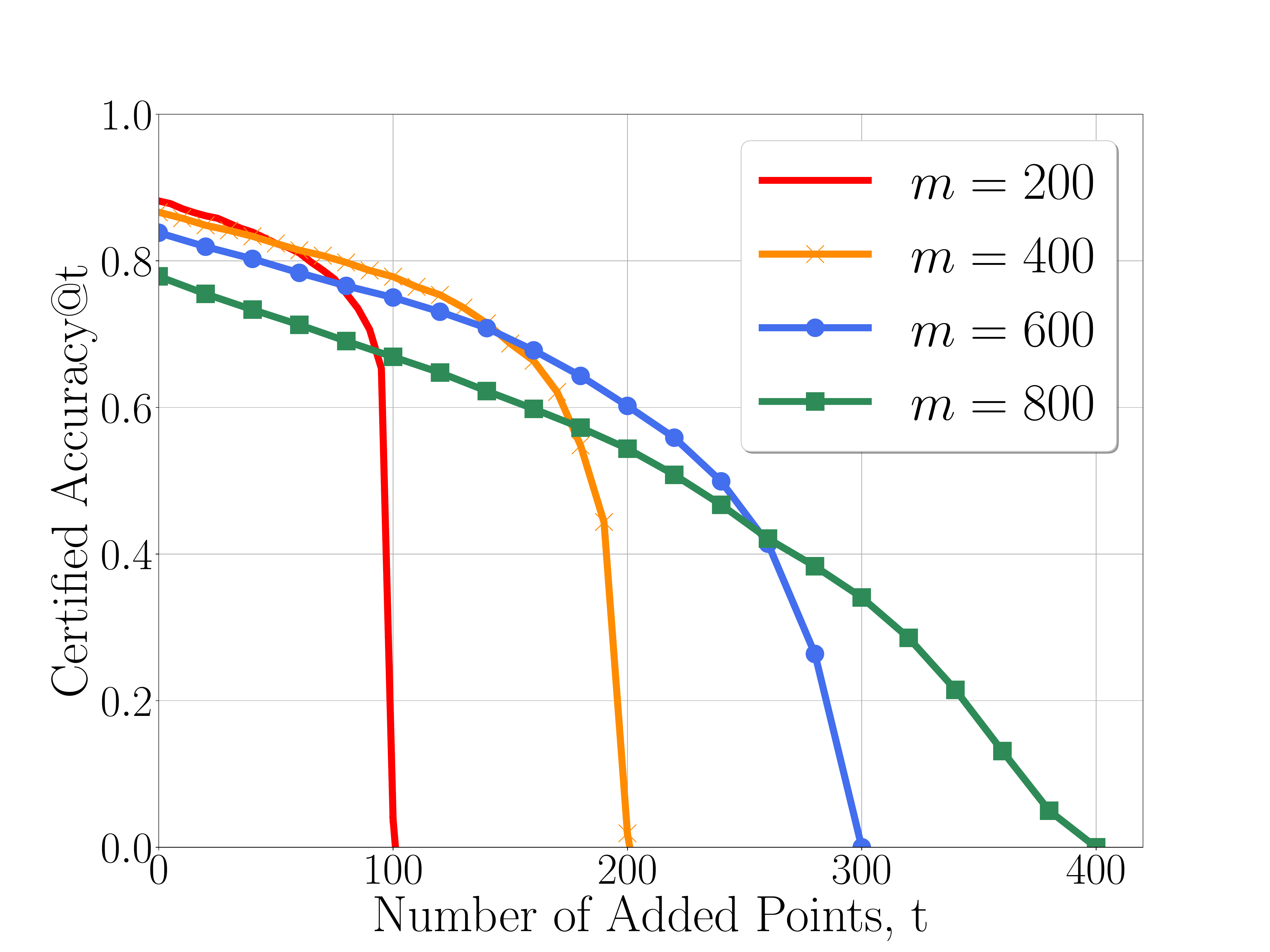}\label{impact_of_N_modelnet40}}   
\subfloat[]{\includegraphics[width =0.23\textwidth]{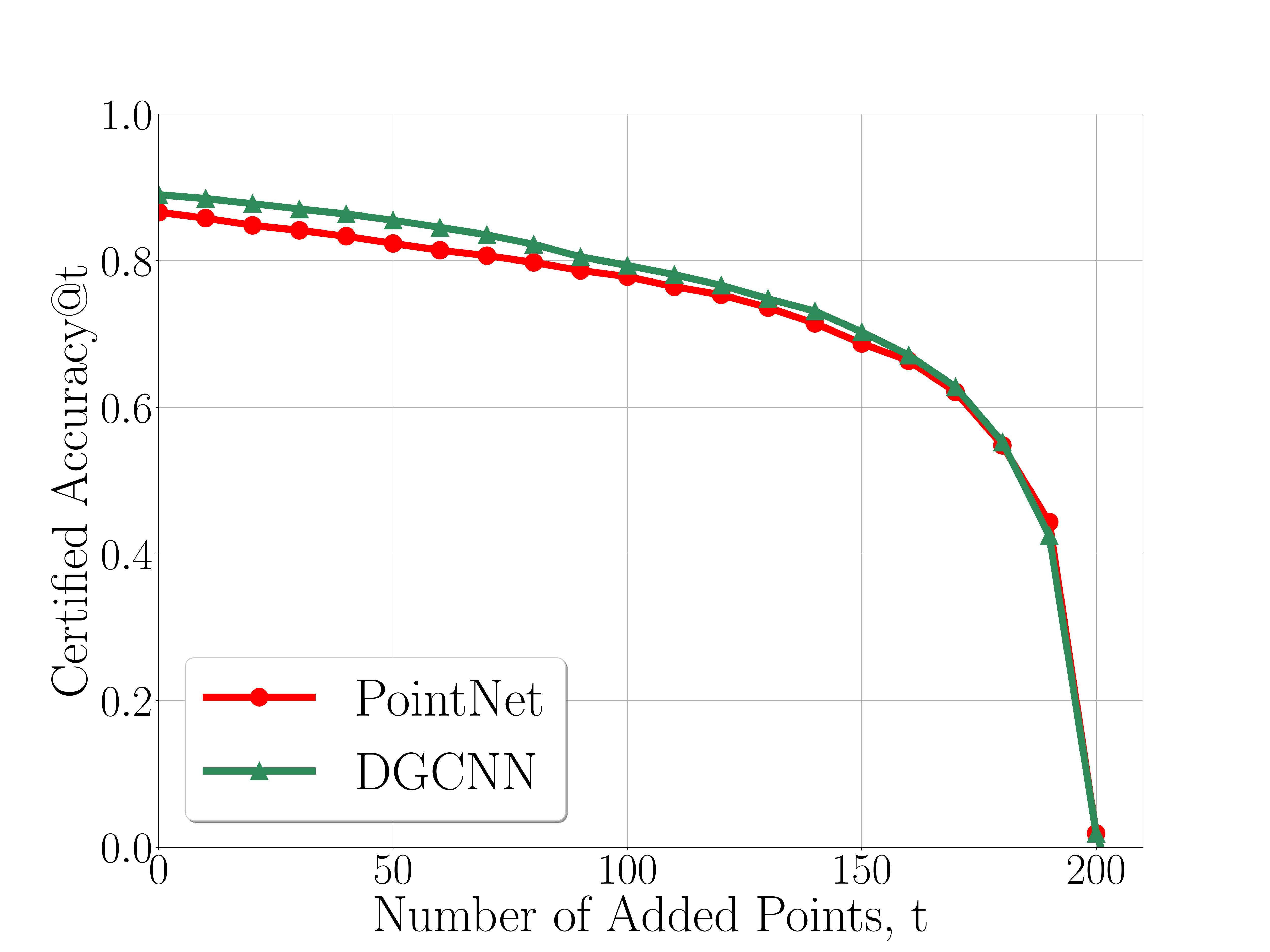}\label{compare_different_base_classifiers}}    
\vspace{-2mm}
\caption{(a) Impact of $m$ on {\name}. (b) Comparing different base point cloud classifiers.  Scenario II is considered.}
\label{fig:group_num}
\vspace{-5mm}
\end{figure}

\subsubsection{Scenario II}
\vspace{-2mm}
\myparatight{Impact of $m$} Figure~\ref{impact_of_N_modelnet40} shows the impact of $m$ on certified accuracy of {\name}. As the results show, $m$ achieves a tradeoff between  accuracy without attacks (i.e., $t=0$) and robustness. In particular, when $m$ is smaller, {\name} can achieve a higher   accuracy without attacks, but is less robust (i.e., certified accuracy drops to 0 more quickly). The reason is that a smaller $m$ means each sub-point cloud includes more points and thus is more likely to be classified correctly, but each adversarially perturbed point impacts a larger fraction of the $m$ sub-point clouds. 

\myparatight{Impact of different base point cloud classifiers} Figure~\ref{compare_different_base_classifiers} compares the certified accuracy of {\name} for different base point cloud classifiers. The experimental results demonstrate that {\name} achieves nearly similar certified accuracy with different base point cloud classifiers. 

\begin{figure}[!t]
    \centering
    \subfloat[]{\includegraphics[width =0.23\textwidth]{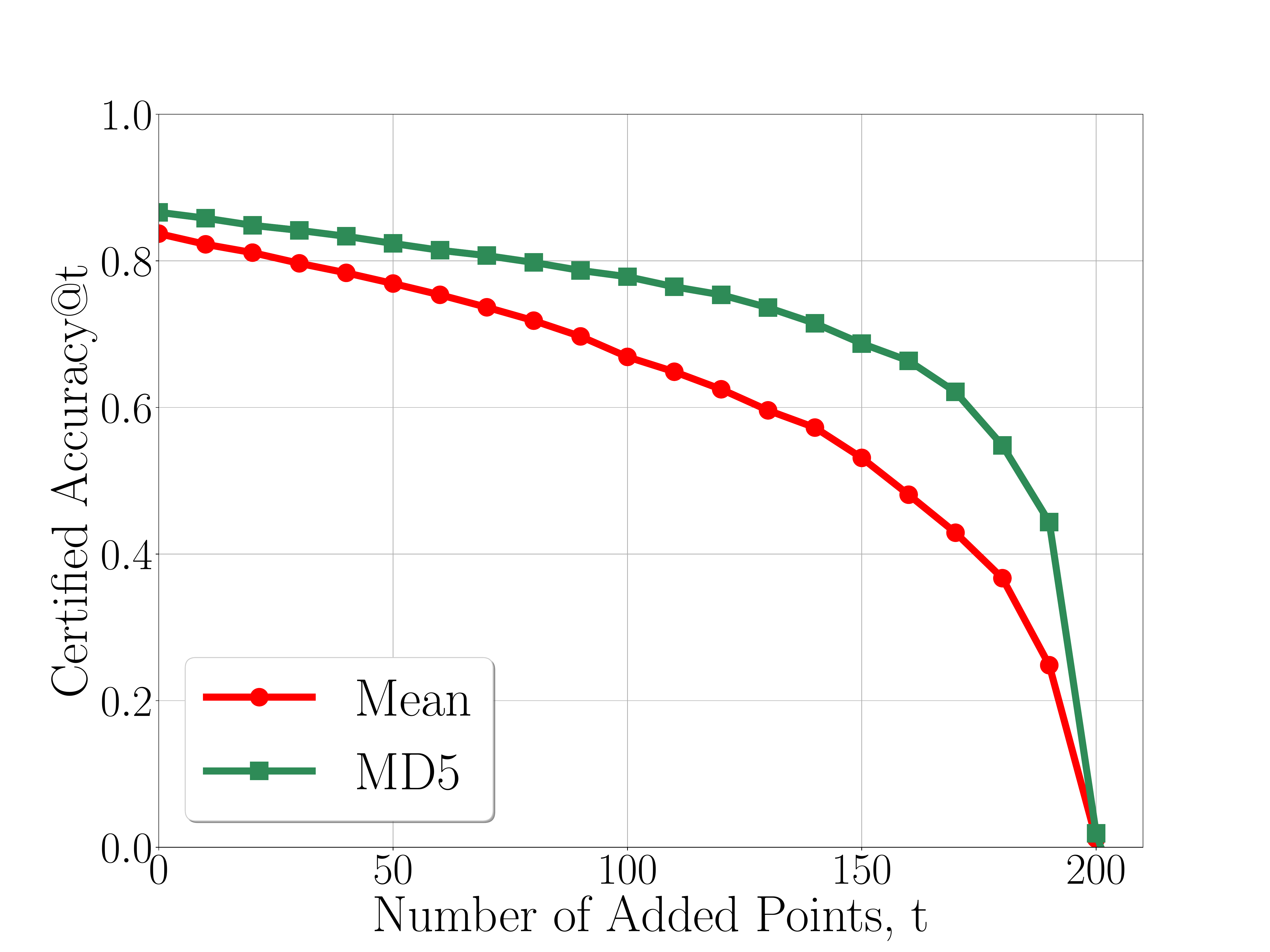}\label{fig:compare_hash_certified_accuracy}}
    \subfloat[]{\includegraphics[width =0.23\textwidth]{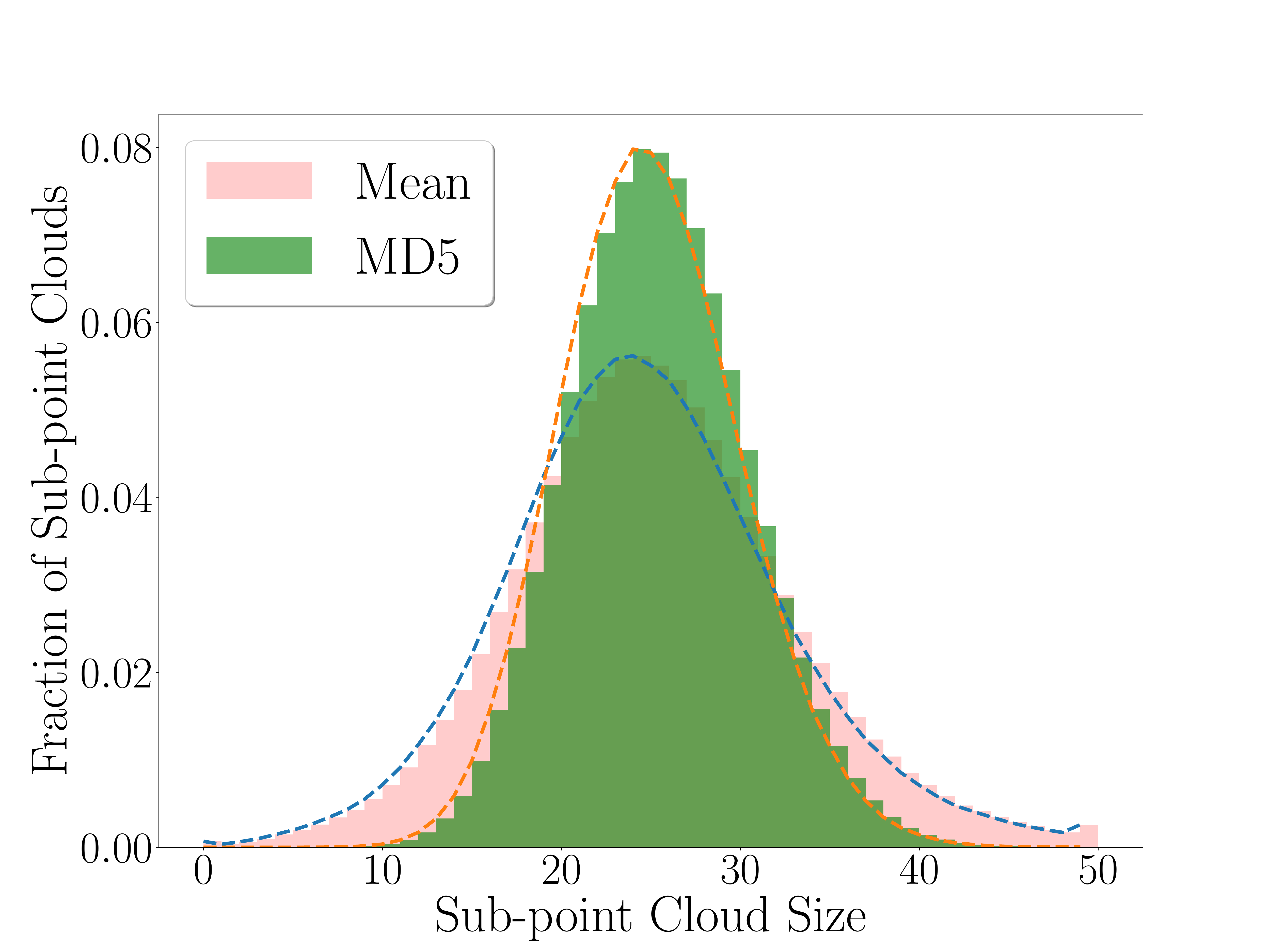}\label{fig:compare_hash_distribution_of_number-of_points}}
    \vspace{-2mm}
    \caption{ (a) Comparing the certified accuracy of {\name} with two hash functions in Scenario II. (b) The distribution of the sub-point cloud sizes for the two hash functions.}
    \label{fig:hash-method}
    \vspace{-2mm}
\end{figure}

\myparatight{Impact of hash function} {\name} uses a hash function to divide a point cloud into sub-point clouds. We compare the cryptographic hash function MD5 with a \emph{mean} based one. In the mean based hash function, given a point $\mathbf{e}_i$, we first compute the mean value of the coordinates of the point (i.e., $
\frac{1}{o}\sum_{j=1,2,\cdots,o}e_{ij}$), then take the first four digits (denoted as $d_1, d_2, d_3, d_4$) of the mean, and finally assign the point $\mathbf{e}_i$ to the $r_i$th sub-point cloud, where $r_i = \sum_{j=1}^{4}d_j\cdot 10^{4-j} \mod m$.  Figure~\ref{fig:compare_hash_certified_accuracy} compares the certified accuracy of {\name} with the two hash functions. Our result indicates that {\name}  achieves higher certified accuracy when using MD5. This is because MD5 generates sub-point clouds with more similar sizes. In particular,  
Figure~\ref{fig:compare_hash_distribution_of_number-of_points} shows the distribution of the number of points in sub-point clouds for the two hash functions. We observe that the  sub-point cloud sizes in MD5 are more concentrated than those in mean. The reason is that cryptographic hash function aims to produce uniformly random hash values in its output space. 

\vspace{-3mm}
\subsubsection{Scenario III}
\begin{figure}[!t]
    \centering
  \subfloat[]{\includegraphics[width =0.23\textwidth]{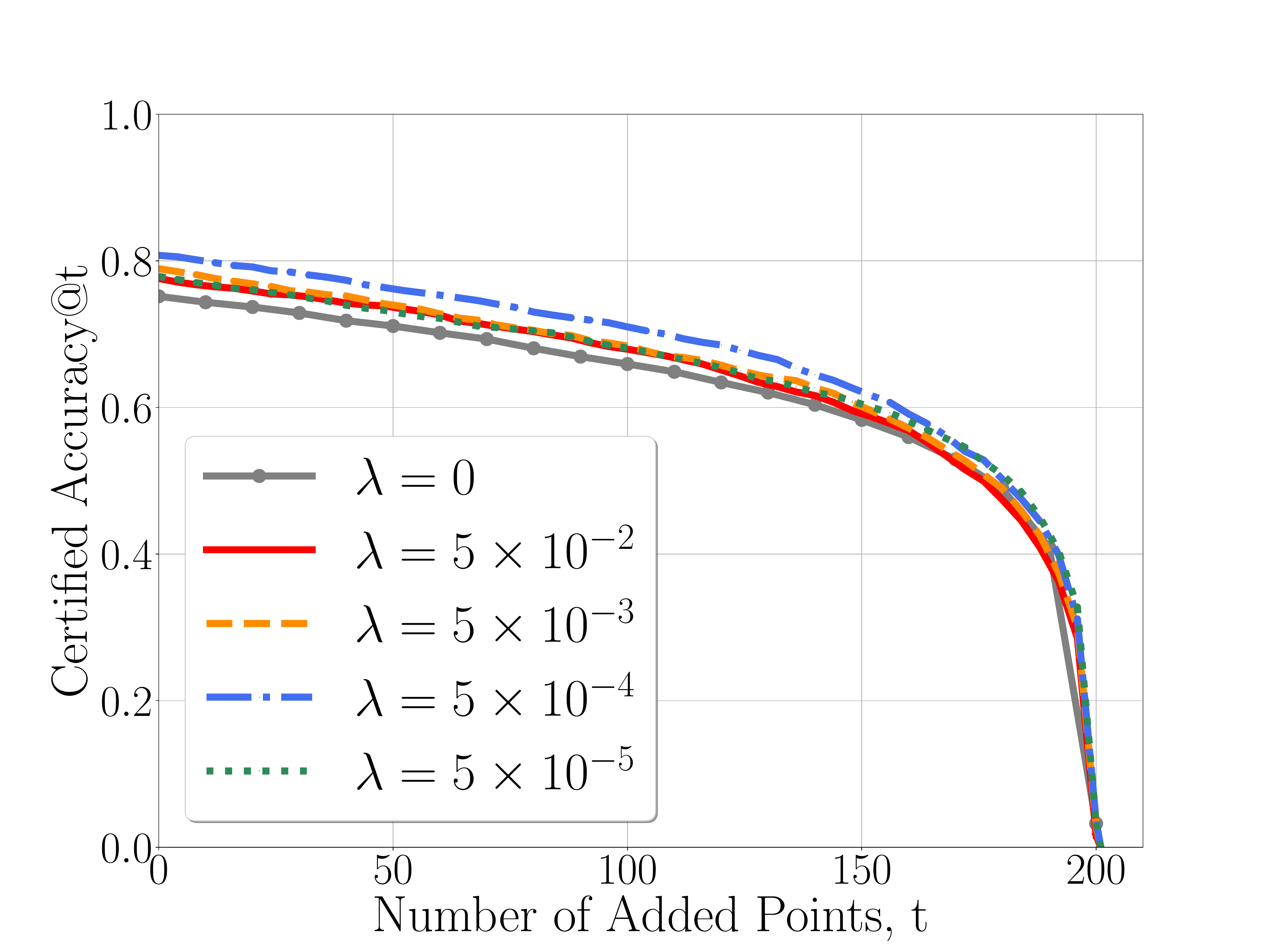}\label{impact_of_lambda_modelnet40}}
     \subfloat[]{\includegraphics[width =0.23\textwidth]{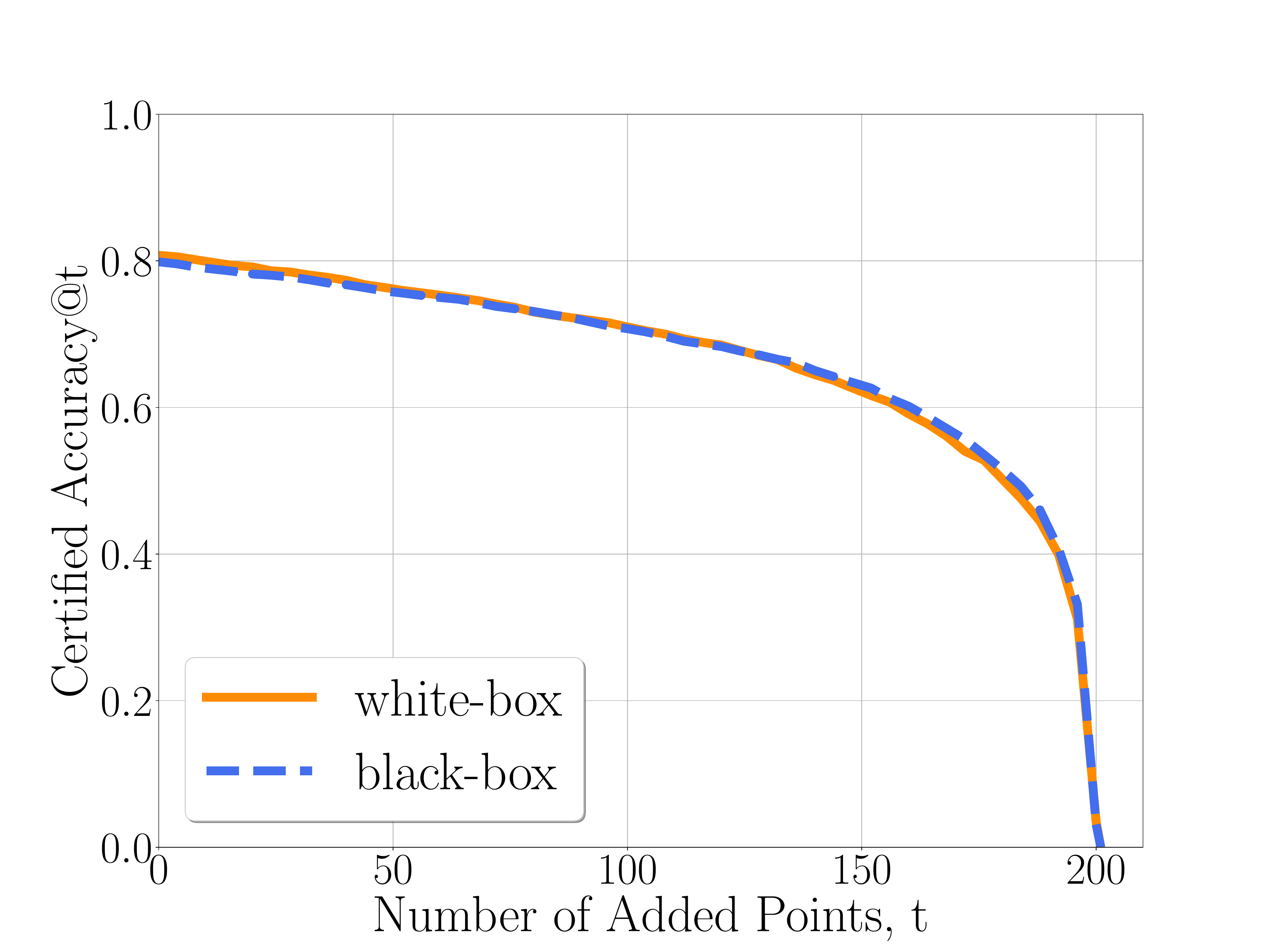}\label{compare_white_black_box_access}}
\vspace{-2mm}
    \caption{ (a) Impact of $\lambda$ on {\name} in Scenario III. (b) Comparing certified accuracy of {\name} in the white-box and black-box settings in Scenario III.}
    \label{fig:ratio}
    \vspace{-5mm}
\end{figure}

\noindent{\bf Impact of $\lambda$:} Figure~\ref{impact_of_lambda_modelnet40} shows the impact of $\lambda$. The certified accuracy first increases  and then decreases as $\lambda$ increases. The reasons are as follows. When $\lambda$ is too small, the base point cloud classifier is less accurate in classifying the point clouds completed by the PCN. When $\lambda$ is too large,  the PCN  completes point clouds with low fidelity, as shown in Figure~\ref{fig:lambda_appendix} in Appendix. The fact that $\lambda>0$ outperforms $\lambda=0$ indicates that our new loss term $L_c(\mathcal{D}_l, \mathcal{C}, \baseclf)$ for training PCN improves {\name}.

\noindent{\bf White-box vs. black-box:}  In Scenario III, a customer uses different methods to train a PCN in the white-box and black-box settings. Figure~\ref{compare_white_black_box_access} compares the certified accuracy of {\name} in the two settings. The results show that {\name} can achieve similar certified accuracy in both settings, which means that the distilled student model approximates the teacher model well. We also found that when the student model and teacher model have different architectures in the black-box setting, our {\name} still achieves high certified accuracy. Due to limited space, we show the results in Figure~\ref{fig:teachermodel} in Appendix. 

\begin{figure}
    \centering
   \subfloat[]{\includegraphics[width =0.23\textwidth]{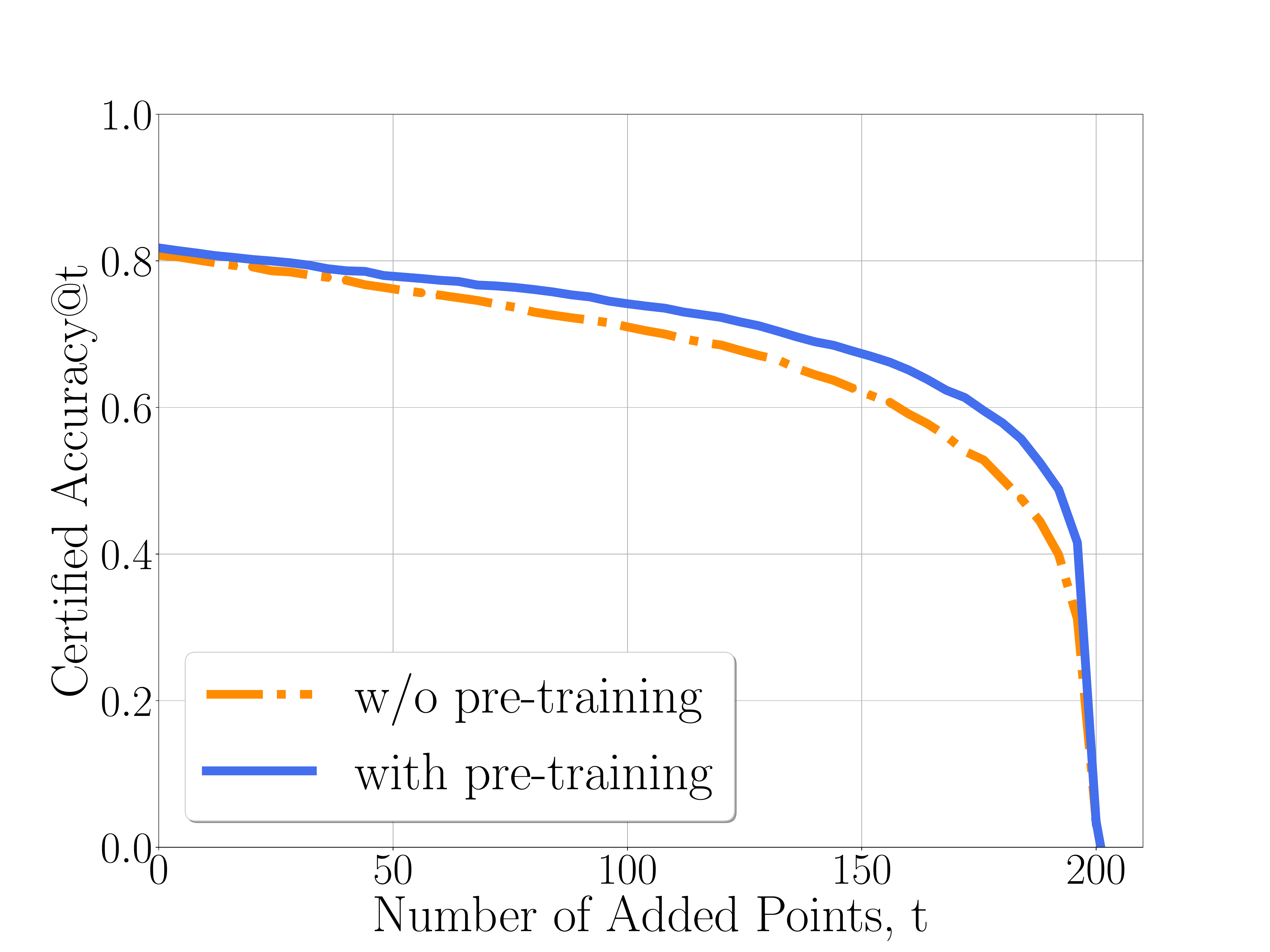}\label{pretrained_improve_certified_modelnet40}} 
  \subfloat[]{\includegraphics[width =0.23\textwidth]{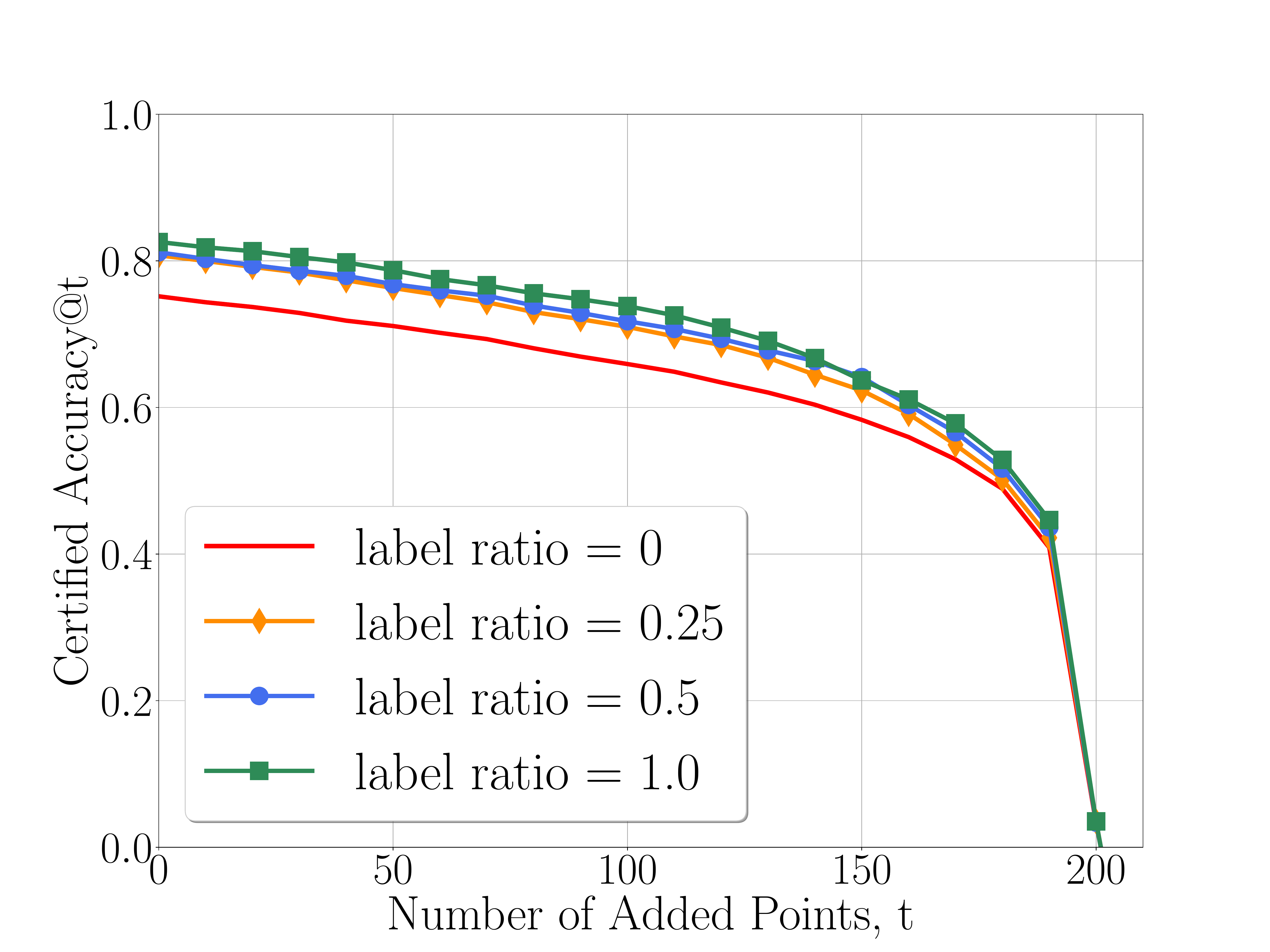}\label{impact_of_ratio_modelnet40}}
    \vspace{-2mm}
    \caption{(a) Pre-trained PCN improves {\name} in Scenario III.  (b) Impact of the fraction of a customer's labeled point clouds on {\name} in Scenario III.}
    \label{fig:pretrained}
    \vspace{-5mm}
\end{figure}

\noindent{\bf Pre-trained PCN improves certified accuracy:} In our previous experiments, we assume a customer trains a PCN from scratch. However, when a customer has a small amount of point clouds, it may be hard to train a good PCN from scratch. {To address the issue, the customer could fine-tune a pre-trained PCN instead of training from scratch. We pretrain a PCN using 8-class of ShapeNet \cite{chang2015shapenet} (used in \cite{liu2020morphing}) and adopt it in our experiment.} Figure~\ref{pretrained_improve_certified_modelnet40} shows our results, which indicates that pre-trained PCN can improve the certified accuracy of {\name}.

\noindent{\bf Impact of label ratio:} Figure~\ref{impact_of_ratio_modelnet40} shows the impact of the fraction of a customer's point clouds that are labeled  on the certified accuracy of {\name}. We observe {\name}  achieves higher certified accuracy when a customer has more labeled point clouds. This is because with more labeled point clouds, the learnt PCN outputs completed point clouds that are classified by the base point cloud classifier with a higher accuracy. 

\vspace{-3mm}
\section{Conclusion}
\vspace{-1mm}
In this paper, we propose the first certified defense, namely {\name}, that has deterministic robustness guarantees against point addition (or deletion or modification or perturbation) attacks to point cloud classification. Moreover, we propose methods to optimize the performance of {\name} in multiple  application scenarios. Interesting future work includes: 1) exploring/designing hash functions to further improve the robustness guarantees of {\name}, and 2) generalizing {\name} to other domains, e.g., graph.

\myparatight{Acknowledgements}
We thank the anonymous reviewers for their constructive comments.  This work was supported by NSF under grant No. 2112562, 1937786, and 1937787, ARO grant No. W911NF2110182, and Facebook Research Award.

{\small
\bibliographystyle{ieee_fullname}
\bibliography{PointCert}

\begin{thebibliography}{10}\itemsep=-1pt

\bibitem{modelnet40}
{ModelNet40}.
\newblock \url{https://modelnet.cs.princeton.edu/}, 2015.
\newblock Accessed: 2022-06.

\bibitem{scanobjectnn}
{ScanObjectNN}.
\newblock \url{https://github.com/hkust-vgd/scanobjectnn}, 2019.
\newblock Accessed: 2022-06.

\bibitem{cao2017mitigating}
Xiaoyu Cao and Neil~Zhenqiang Gong.
\newblock Mitigating evasion attacks to deep neural networks via region-based
  classification.
\newblock In {\em Annual Computer Security Applications Conference}, 2017.

\bibitem{chang2015shapenet}
Angel~X Chang, Thomas Funkhouser, Leonidas Guibas, Pat Hanrahan, Qixing Huang,
  Zimo Li, Silvio Savarese, Manolis Savva, Shuran Song, Hao Su, et~al.
\newblock Shapenet: An information-rich 3d model repository.
\newblock {\em arXiv preprint arXiv:1512.03012}, 2015.

\bibitem{cohen2019certified}
Jeremy Cohen, Elan Rosenfeld, and Zico Kolter.
\newblock Certified adversarial robustness via randomized smoothing.
\newblock In {\em International Conference on Machine Learning}, 2019.

\bibitem{denipitiyage2021provable}
Dishanika~Dewani Denipitiyage, Thalaiyasingam Ajanthan, Parameswaran
  Kamalaruban, and Adrian Weller.
\newblock Provable defense against clustering attacks on 3d point clouds.
\newblock In {\em The AAAI-22 Workshop on Adversarial Machine Learning and
  Beyond}, 2021.

\bibitem{dong2020self}
Xiaoyi Dong, Dongdong Chen, Hang Zhou, Gang Hua, Weiming Zhang, and Nenghai Yu.
\newblock Self-robust 3d point recognition via gather-vector guidance.
\newblock In {\em IEEE/CVF Conference on Computer Vision and Pattern
  Recognition}, 2020.

\bibitem{fan2017point}
Haoqiang Fan, Hao Su, and Leonidas~J Guibas.
\newblock A point set generation network for 3d object reconstruction from a
  single image.
\newblock In {\em IEEE/CVF Conference on Computer Vision and Pattern
  Recognition}, 2017.

\bibitem{fischer2021scalable}
Marc Fischer, Maximilian Baader, and Martin Vechev.
\newblock Scalable certified segmentation via randomized smoothing.
\newblock In {\em International Conference on Machine Learning}, 2021.

\bibitem{ghadimi2013stochastic}
Saeed Ghadimi and Guanghui Lan.
\newblock Stochastic first-and zeroth-order methods for nonconvex stochastic
  programming.
\newblock {\em SIAM Journal on Optimization}, 2013.

\bibitem{hamdi2021mvtn}
Abdullah Hamdi, Silvio Giancola, and Bernard Ghanem.
\newblock Mvtn: Multi-view transformation network for 3d shape recognition.
\newblock In {\em IEEE/CVF International Conference on Computer Vision}, 2021.

\bibitem{hamdi2020advpc}
Abdullah Hamdi, Sara Rojas, Ali Thabet, and Bernard Ghanem.
\newblock Advpc: Transferable adversarial perturbations on 3d point clouds.
\newblock In {\em European Conference on Computer Vision}, 2020.

\bibitem{hinton2015distilling}
Geoffrey Hinton, Oriol Vinyals, and Jeff Dean.
\newblock Distilling the knowledge in a neural network.
\newblock {\em arXiv preprint arXiv:1503.02531}, 2015.

\bibitem{huang2020pf}
Zitian Huang, Yikuan Yu, Jiawen Xu, Feng Ni, and Xinyi Le.
\newblock Pf-net: Point fractal network for 3d point cloud completion.
\newblock In {\em IEEE/CVF Conference on Computer Vision and Pattern
  Recognition}, 2020.

\bibitem{jia2022almost}
Jinyuan Jia, Binghui Wang, Xiaoyu Cao, Hongbin Liu, and Neil~Zhenqiang Gong.
\newblock Almost tight l0-norm certified robustness of top-k predictions
  against adversarial perturbations.
\newblock In {\em International Conference on Learning Representations}, 2022.

\bibitem{kim2021minimal}
Jaeyeon Kim, Binh-Son Hua, Thanh Nguyen, and Sai-Kit Yeung.
\newblock Minimal adversarial examples for deep learning on 3d point clouds.
\newblock In {\em IEEE/CVF International Conference on Computer Vision}, 2021.

\bibitem{lecuyer2019certified}
Mathias Lecuyer, Vaggelis Atlidakis, Roxana Geambasu, Daniel Hsu, and Suman
  Jana.
\newblock Certified robustness to adversarial examples with differential
  privacy.
\newblock In {\em IEEE Symposium on Security and Privacy}, 2019.

\bibitem{li2018second}
Bai Li, Changyou Chen, Wenlin Wang, and Lawrence Carin.
\newblock Second-order adversarial attack and certifiable robustness.
\newblock 2018.

\bibitem{li2022robust}
Kaidong Li, Ziming Zhang, Cuncong Zhong, and Guanghui Wang.
\newblock Robust structured declarative classifiers for 3d point clouds:
  Defending adversarial attacks with implicit gradients.
\newblock In {\em IEEE/CVF Conference on Computer Vision and Pattern
  Recognition}, 2022.

\bibitem{liu2022imperceptible}
Daizong Liu and Wei Hu.
\newblock Imperceptible transfer attack and defense on 3d point cloud
  classification.
\newblock {\em IEEE Transactions on Pattern Analysis and Machine Intelligence},
  2022.

\bibitem{liu2019extending}
Daniel Liu, Ronald Yu, and Hao Su.
\newblock Extending adversarial attacks and defenses to deep 3d point cloud
  classifiers.
\newblock In {\em IEEE International Conference on Image Processing}, 2019.

\bibitem{liu2020adversarial}
Daniel Liu, Ronald Yu, and Hao Su.
\newblock Adversarial shape perturbations on 3d point clouds.
\newblock In {\em European Conference on Computer Vision}, 2020.

\bibitem{liu2021pointguard}
Hongbin Liu, Jinyuan Jia, and Neil~Zhenqiang Gong.
\newblock Pointguard: Provably robust 3d point cloud classification.
\newblock In {\em IEEE/CVF Conference on Computer Vision and Pattern
  Recognition}, 2021.

\bibitem{liu2020morphing}
Minghua Liu, Lu Sheng, Sheng Yang, Jing Shao, and Shi-Min Hu.
\newblock Morphing and sampling network for dense point cloud completion.
\newblock In {\em AAAI Conference on Artificial Intelligence}, 2020.

\bibitem{liu2018towards}
Xuanqing Liu, Minhao Cheng, Huan Zhang, and Cho-Jui Hsieh.
\newblock Towards robust neural networks via random self-ensemble.
\newblock In {\em European Conference on Computer Vision}, 2018.

\bibitem{lorenz2021robustness}
Tobias Lorenz, Anian Ruoss, Mislav Balunovi{\'c}, Gagandeep Singh, and Martin
  Vechev.
\newblock Robustness certification for point cloud models.
\newblock In {\em IEEE/CVF International Conference on Computer Vision}, 2021.

\bibitem{ma2020efficient}
Chengcheng Ma, Weiliang Meng, Baoyuan Wu, Shibiao Xu, and Xiaopeng Zhang.
\newblock Efficient joint gradient based attack against sor defense for 3d
  point cloud classification.
\newblock In {\em ACM International Conference on Multimedia}, 2020.

\bibitem{perez20223deformrs}
Juan~C P{\'e}rez, Motasem Alfarra, Silvio Giancola, Bernard Ghanem, et~al.
\newblock 3deformrs: Certifying spatial deformations on point clouds.
\newblock In {\em IEEE/CVF Conference on Computer Vision and Pattern
  Recognition}, 2022.

\bibitem{qi2017pointnet}
Charles~R Qi, Hao Su, Kaichun Mo, and Leonidas~J Guibas.
\newblock Pointnet: Deep learning on point sets for 3d classification and
  segmentation.
\newblock In {\em IEEE/CVF Conference on Computer Vision and Pattern
  Recognition}, 2017.

\bibitem{qiu2021geometric}
Shi Qiu, Saeed Anwar, and Nick Barnes.
\newblock Geometric back-projection network for point cloud classification.
\newblock {\em IEEE Transactions on Multimedia}, 2021.

\bibitem{salman2019provably}
Hadi Salman, Jerry Li, Ilya Razenshteyn, Pengchuan Zhang, Huan Zhang, Sebastien
  Bubeck, and Greg Yang.
\newblock Provably robust deep learning via adversarially trained smoothed
  classifiers.
\newblock {\em Advances in Neural Information Processing Systems}, 2019.

\bibitem{shen2021interpreting}
Wen Shen, Qihan Ren, Dongrui Liu, and Quanshi Zhang.
\newblock Interpreting representation quality of dnns for 3d point cloud
  processing.
\newblock {\em Advances in Neural Information Processing Systems}, 2021.

\bibitem{sun2021adversarially}
Jiachen Sun, Yulong Cao, Christopher~B Choy, Zhiding Yu, Anima Anandkumar,
  Zhuoqing~Morley Mao, and Chaowei Xiao.
\newblock Adversarially robust 3d point cloud recognition using
  self-supervisions.
\newblock {\em Advances in Neural Information Processing Systems}, 2021.

\bibitem{sun2020adversarial}
Jiachen Sun, Karl Koenig, Yulong Cao, Qi~Alfred Chen, and Z~Morley Mao.
\newblock On adversarial robustness of 3d point cloud classification under
  adaptive attacks.
\newblock {\em arXiv preprint arXiv:2011.11922}, 2020.

\bibitem{sun2022benchmarking}
Jiachen Sun, Qingzhao Zhang, Bhavya Kailkhura, Zhiding Yu, Chaowei Xiao, and
  Z~Morley Mao.
\newblock Benchmarking robustness of 3d point cloud recognition against common
  corruptions.
\newblock {\em arXiv preprint arXiv:2201.12296}, 2022.

\bibitem{uy2019revisiting}
Mikaela~Angelina Uy, Quang-Hieu Pham, Binh-Son Hua, Thanh Nguyen, and Sai-Kit
  Yeung.
\newblock Revisiting point cloud classification: A new benchmark dataset and
  classification model on real-world data.
\newblock In {\em IEEE/CVF International Conference on Computer Vision}, 2019.

\bibitem{wang2020cascaded}
Xiaogang Wang, Marcelo~H Ang~Jr, and Gim~Hee Lee.
\newblock Cascaded refinement network for point cloud completion.
\newblock In {\em IEEE/CVF Conference on Computer Vision and Pattern
  Recognition}, 2020.

\bibitem{wang2019dynamic}
Yue Wang, Yongbin Sun, Ziwei Liu, Sanjay~E Sarma, Michael~M Bronstein, and
  Justin~M Solomon.
\newblock Dynamic graph cnn for learning on point clouds.
\newblock {\em ACM Transactions on Graphics (TOG)}, 2019.

\bibitem{wen2020point}
Xin Wen, Tianyang Li, Zhizhong Han, and Yu-Shen Liu.
\newblock Point cloud completion by skip-attention network with hierarchical
  folding.
\newblock In {\em IEEE/CVF Conference on Computer Vision and Pattern
  Recognition}, 2020.

\bibitem{wicker2019robustness}
Matthew Wicker and Marta Kwiatkowska.
\newblock Robustness of 3d deep learning in an adversarial setting.
\newblock In {\em IEEE/CVF Conference on Computer Vision and Pattern
  Recognition}, 2019.

\bibitem{wu2020if}
Ziyi Wu, Yueqi Duan, He Wang, Qingnan Fan, and Leonidas~J Guibas.
\newblock If-defense: 3d adversarial point cloud defense via implicit function
  based restoration.
\newblock {\em arXiv preprint arXiv:2010.05272}, 2020.

\bibitem{wu20153d}
Zhirong Wu, Shuran Song, Aditya Khosla, Fisher Yu, Linguang Zhang, Xiaoou Tang,
  and Jianxiong Xiao.
\newblock 3d shapenets: A deep representation for volumetric shapes.
\newblock In {\em IEEE/CVF Conference on Computer Vision and Pattern
  Recognition}, 2015.

\bibitem{xiang2019generating}
Chong Xiang, Charles~R Qi, and Bo Li.
\newblock Generating 3d adversarial point clouds.
\newblock In {\em IEEE/CVF Conference on Computer Vision and Pattern
  Recognition}, 2019.

\bibitem{xiang2021walk}
Tiange Xiang, Chaoyi Zhang, Yang Song, Jianhui Yu, and Weidong Cai.
\newblock Walk in the cloud: Learning curves for point clouds shape analysis.
\newblock {\em arXiv preprint arXiv:2105.01288}, 2021.

\bibitem{xie2020grnet}
Haozhe Xie, Hongxun Yao, Shangchen Zhou, Jiageng Mao, Shengping Zhang, and
  Wenxiu Sun.
\newblock Grnet: Gridding residual network for dense point cloud completion.
\newblock In {\em European Conference on Computer Vision}, 2020.

\bibitem{yang2019adversarial}
Jiancheng Yang, Qiang Zhang, Rongyao Fang, Bingbing Ni, Jinxian Liu, and Qi
  Tian.
\newblock Adversarial attack and defense on point sets.
\newblock {\em arXiv preprint arXiv:1902.10899}, 2019.

\bibitem{yu2021pointr}
Xumin Yu, Yongming Rao, Ziyi Wang, Zuyan Liu, Jiwen Lu, and Jie Zhou.
\newblock Pointr: Diverse point cloud completion with geometry-aware
  transformers.
\newblock In {\em IEEE/CVF International Conference on Computer Vision}, 2021.

\bibitem{yuan2018pcn}
Wentao Yuan, Tejas Khot, David Held, Christoph Mertz, and Martial Hebert.
\newblock Pcn: Point completion network.
\newblock In {\em International Conference on 3D Vision (3DV)}, 2018.

\bibitem{zhang2022pointcutmix}
Jinlai Zhang, Lyujie Chen, Bo Ouyang, Binbin Liu, Jihong Zhu, Yujin Chen,
  Yanmei Meng, and Danfeng Wu.
\newblock Pointcutmix: Regularization strategy for point cloud classification.
\newblock {\em Neurocomputing}, 2022.

\bibitem{zhao2021point}
Hengshuang Zhao, Li Jiang, Jiaya Jia, Philip~HS Torr, and Vladlen Koltun.
\newblock Point transformer.
\newblock In {\em IEEE/CVF International Conference on Computer Vision}, 2021.

\bibitem{zhao2020isometry}
Yue Zhao, Yuwei Wu, Caihua Chen, and Andrew Lim.
\newblock On isometry robustness of deep 3d point cloud models under
  adversarial attacks.
\newblock In {\em IEEE/CVF Conference on Computer Vision and Pattern
  Recognition}, 2020.

\bibitem{zheng2019pointcloud}
Tianhang Zheng, Changyou Chen, Junsong Yuan, Bo Li, and Kui Ren.
\newblock Pointcloud saliency maps.
\newblock In {\em IEEE/CVF International Conference on Computer Vision}, 2019.

\bibitem{zhou2019dup}
Hang Zhou, Kejiang Chen, Weiming Zhang, Han Fang, Wenbo Zhou, and Nenghai Yu.
\newblock Dup-net: Denoiser and upsampler network for 3d adversarial point
  clouds defense.
\newblock In {\em IEEE/CVF International Conference on Computer Vision}, 2019.

\end{thebibliography}
}

\appendix
\newpage
\newpage

\section{Proof of Theorem~\ref{tightnesstheorem}}
\label{proof_of_theorem2}
We prove the theorem by constructing a base point cloud classifier $\baseclf'$ and an adversarial point cloud $\pointcloud'$. 
Suppose $\pointcloud_1, \pointcloud_2, \cdots, \pointcloud_m$ are the $m$ sub-point clouds for the point cloud $\pointcloud$. We let $\baseclf'$ predict label $y$ for $\pointcloud_j$, where $j=1,2,\cdots, M_y(\pointcloud)$, and predict label $l$ ($l\neq y$) for $M_l(\pointcloud)$ sub-point clouds among the remaining $m-M_y(\pointcloud)$ sub-point clouds. When an attacker can arbitrarily add (or delete or modify) at most $t'(\pointcloud)+1$ points to $\pointcloud$, we construct the following $\pointcloud'$. For point addition (or deletion) attacks, we can find $\pointcloud'$ such that a point is added (or deleted) to $\pointcloud_j$, where $j=1,2,\cdots,t'(\pointcloud)+1$. For simplicity, we use $\pointcloud'_j$ to denote the corresponding sub-point cloud by adding (or deleting) a point to $\pointcloud_j$. For point modification/perturbation attacks, we can find $\pointcloud'$ such that a point is deleted from $\pointcloud_{2\cdot j-1}$ and a point is added to $\pointcloud_{2\cdot j}$, where $j=1,2,\cdots,t'(\pointcloud)+1$. For simplicity, we use $\pointcloud'_{2\cdot j-1}$ and $\pointcloud'_{2\cdot j}$ to denote the corresponding sub-point clouds. Suppose $l' = \argmax_{l\neq y} (M_l(\pointcloud)+\mathbb{I}(y>l))$. We let $\baseclf'$ predict label $l'$ for sub-point clouds $\pointcloud'_j$, where $j=1,2,\cdots, \tau \cdot (t'(\pointcloud)+1)$ and $\tau$ is 1 (or 1 or 2 or 2) for point addition (or deletion or modification or perturbation) attacks. Given the constructed $f'$ and $\pointcloud'$, we have $M_y(\pointcloud')=M_y(\pointcloud)-\tau \cdot (t'(\pointcloud)+1)$ and $M_{l'}(\pointcloud')=M_{l'}(\pointcloud)+\tau \cdot (t'(\pointcloud)+1)$. Then, we have the following:
\begin{align}
& M_{l'}(\pointcloud') + \mathbb{I}(y>l') \\
=&M_{l'}(\pointcloud)+\tau \cdot (t'(\pointcloud)+1)+ \mathbb{I}(y>l') \\
\label{tight_proof_key_step_1}
=&M_{l'}(\pointcloud)+(2\cdot\tau -\tau) \cdot (t'(\pointcloud)+1)+ \mathbb{I}(y>l') \\
> & M_{l'}(\pointcloud) +2\cdot \tau \cdot \frac{M_y(\pointcloud)- (M_{l'}(\pointcloud)+\mathbb{I}(y>l'))}{2\cdot \tau} \nonumber\\
\label{tight_proof_key_step_2}
&-\tau \cdot (t'(\pointcloud)+1)+ \mathbb{I}(y>l') \\
=&M_y(\pointcloud) -\tau \cdot (t'(\pointcloud)+1) \\
=& M_y(\pointcloud').
\end{align}
We have Equation~(\ref{tight_proof_key_step_2}) from Equation~(\ref{tight_proof_key_step_1}) based on the fact that $\lfloor x \rfloor +1 > x$, where $\lfloor \cdot \rfloor$ is floor function and $x$ is an arbitrary non-negative real number. 
Therefore, the ensemble point cloud classifier $\ensclf'$ built upon $\baseclf'$ predicts label $l'$ instead of $y$ for the constructed point cloud $\pointcloud'$.

\section{Loss Term Proposed by Fan et al.~\cite{fan2017point}}
\label{loss_fan_2018_point}
Fan et al.~\cite{fan2017point} proposed the following loss term $L_p(\mathcal{D}_u, \mathcal{C})$ (Chamfer Distance):
{\small 
\begin{align}
  L_p(\mathcal{D}_u, \mathcal{C}) =  \frac{1}{|\mathcal{D}_u|}& \sum_{(\pointcloud_s,\pointcloud_p)\in \mathcal{D}_u}[ \frac{1}{|\mathcal{C}(\pointcloud_s)|}\sum_{\mathbf{e}_s\in \mathcal{C}(\pointcloud_s)}\min_{\mathbf{e}_p\in\pointcloud_p}\lnorm{\mathbf{e}_s-\mathbf{e}_p}_2 \nonumber \\
  +& \frac{1}{|\pointcloud_p|}\sum_{\mathbf{e}_p\in \pointcloud_p}\min_{\mathbf{e}_s\in \mathcal{C}(\pointcloud_s)}\lnorm{\mathbf{e}_s-\mathbf{e}_p}_2],
\end{align}
}
where $\mathcal{C}(\pointcloud_s)$ is the completed point cloud outputted by $\mathcal{C}$ for the sub-point cloud $\pointcloud_s$, and 
 $\mathbf{e}_s$ (or $\mathbf{e}_p$) is a point in $\mathcal{C}(\pointcloud_s)$ (or $\pointcloud_p$).

\begin{figure}
    \centering
    \includegraphics[width =0.8\linewidth]{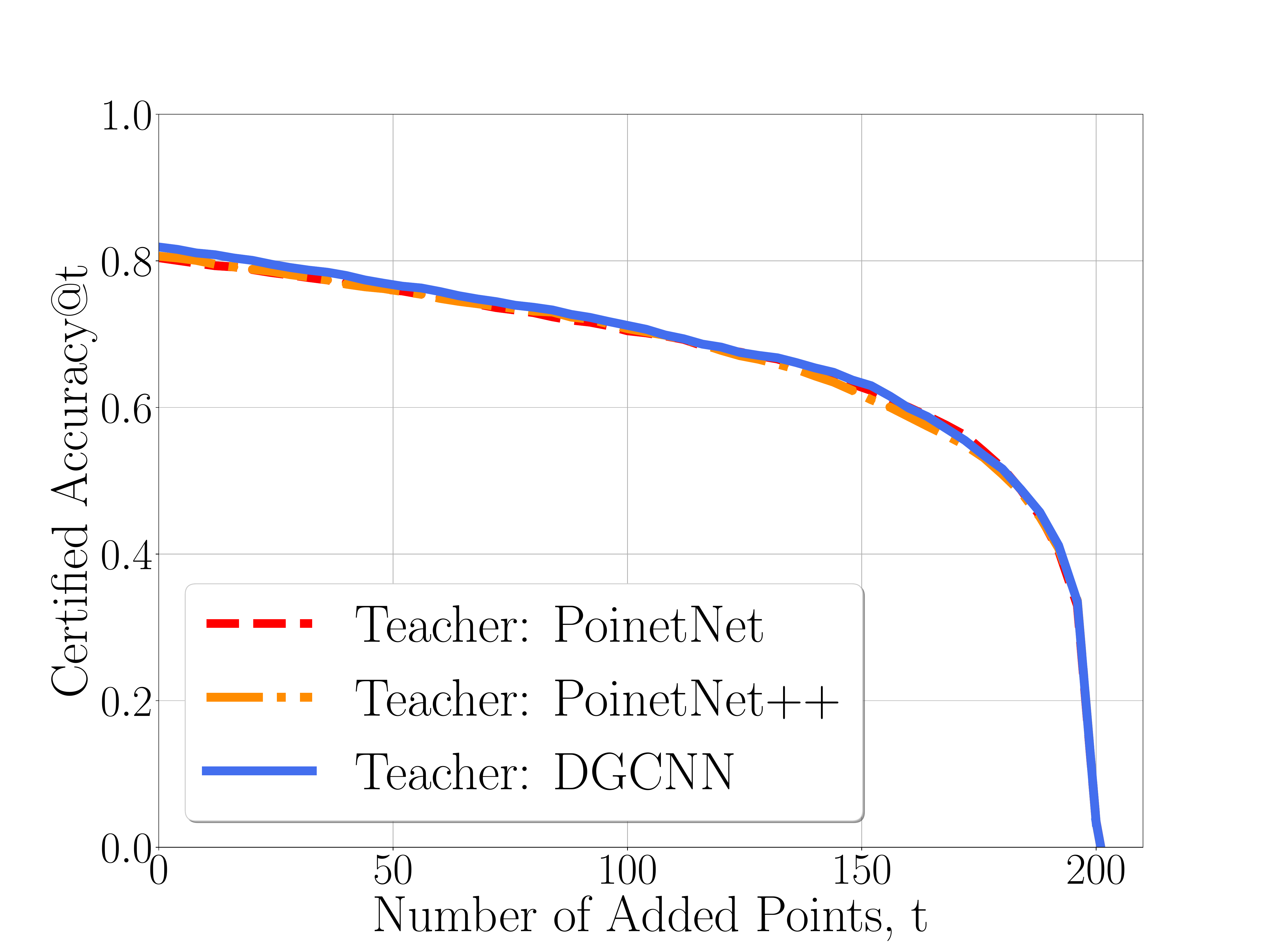}
    \caption{Impact of the teacher model's architecture on the certified accuracy of {\name} in the black-box setting of Scenario III. Student model architecture is PointNet and dataset is ModelNet40.}
    \label{fig:teachermodel}
\end{figure}

\section{Dataset Description}
\label{app:dataset}
We adopt two publicly available benchmark datasets, namely ModelNet40~\cite{wu20153d} and ScanObjectNN~\cite{uy2019revisiting}, in our evaluation. In particular, ModelNet40 contains 9,843 training point clouds and 2,468 testing point clouds. Each point cloud has 10,000 points on average and belongs to one of the 40 categories. In ScanObjectNN dataset, the number of training point clouds and the number of testing point clouds are respectively 2,319 and 583. The total number of classes in this dataset is 15. ScanObjectNN dataset has two variants, namely ScanObjectNN-OBJ\_Only and ScanObjectNN-OBJ\_BG. 
The difference is that the object in ScanObjectNN-OBJ\_Only does not have background while the object in ScanObjectNN-OBJ\_BG has. We use both variants. In accordance with ModelNet40, we keep at most 10,000 points in each point cloud in ScanObjectNN. On average, each point cloud has 9,594 and 9,774 points respectively for the two variants. Under the same setting, we compare with existing defenses and conduct our experiments using raw point clouds to simulate real-world attack scenarios. It is noted that our $\name$ still outperforms previous defenses when the point clouds are all sub-sampled to reduce their sizes as previous defenses~\cite{liu2021pointguard}.

\section{Details of Compared Methods.}
\label{app:method}
We compare {\name} with undefended model, randomized smoothing~\cite{cohen2019certified}, and PointGuard~\cite{liu2021pointguard}.

\begin{itemize}
    \item {\bf Undefended model:} Undefended model is a base point cloud classifier that is trained and tested in the standard way. It does not have certified robustness guarantees.
    
    \item {\bf Randomized smoothing~\cite{cohen2019certified}:} Randomized smoothing builds a certifiably robust classifier via adding a zero-mean Gaussian noise with standard deviation $\sigma$ to an input. In particular, given a testing point cloud,  randomized smoothing constructs $N$ noisy point clouds, each of which is constructed by adding random  Gaussian noise to each dimension of each point of the point cloud. Then,  randomized smoothing uses a point cloud classifier to predict the labels of the noisy point clouds and takes a majority vote among the predicted labels as the final predicted label of the point cloud.  Randomized smoothing provably predicts the same label for a point cloud when the $\ell_2$-norm of the adversarial perturbation added to its points  is less than a threshold (called \emph{certified radius}). 
    
    When applied to point cloud classification, randomized smoothing can only derive certified radius  against point modification attacks~\cite{liu2021pointguard}. Moreover, we can transform certified radius to certified perturbation size via employing the relationship between $\ell_2$-norm and $\ell_0$-norm. In particular, suppose the points in a (adversarial) point cloud lie in a space (denoted as $\Omega$). We assume the largest $\ell_2$-norm distance between two arbitrary points in the space $\Omega$ is bounded by $\eta$. In other words, we have $\eta \geq \max_{\omega_1 \in \Omega, \omega_2 \in \Omega}\lnorm{\omega_1 - \omega_2}_2$. Note that $\eta$ could be different for different datasets. For instance, $\eta$ is respectively $2\sqrt{3}$ and $\sqrt{15}$ on ModelNet40 and ScanObjectNN datasets. Given a certified radius $\gamma$ (under $\ell_2$-norm) obtained by randomized smoothing and the $\eta$, the certified perturbation size can be computed as $\lfloor \gamma^2/\eta^2 \rfloor$. 
 
    \item {\bf PointGuard~\cite{liu2021pointguard}:} PointGuard is the state-of-the-art certified defense against adversarial point clouds. Roughly speaking, given a testing point cloud, PointGuard first creates $N$ subsampled point clouds,  each of which is obtained by randomly subsampling $k$ (a parameter in PointGuard) points from the given point cloud. Then, PointGuard uses a base point cloud classifier to predict labels for those subsampled point clouds. Finally, PointGuard counts the number of subsampled point clouds whose predicted labels are $l$ ($l=1,2,\cdots, c$). The label with the largest count is viewed as the predicted label for the given point cloud. PointGuard provably predicts the same label for a point cloud when the number of arbitrarily added, deleted, and/or modified points is less than a threshold, which is the certified perturbation size.
\end{itemize}

\begin{figure}[!t]
    \centering
    \includegraphics[width =\linewidth]{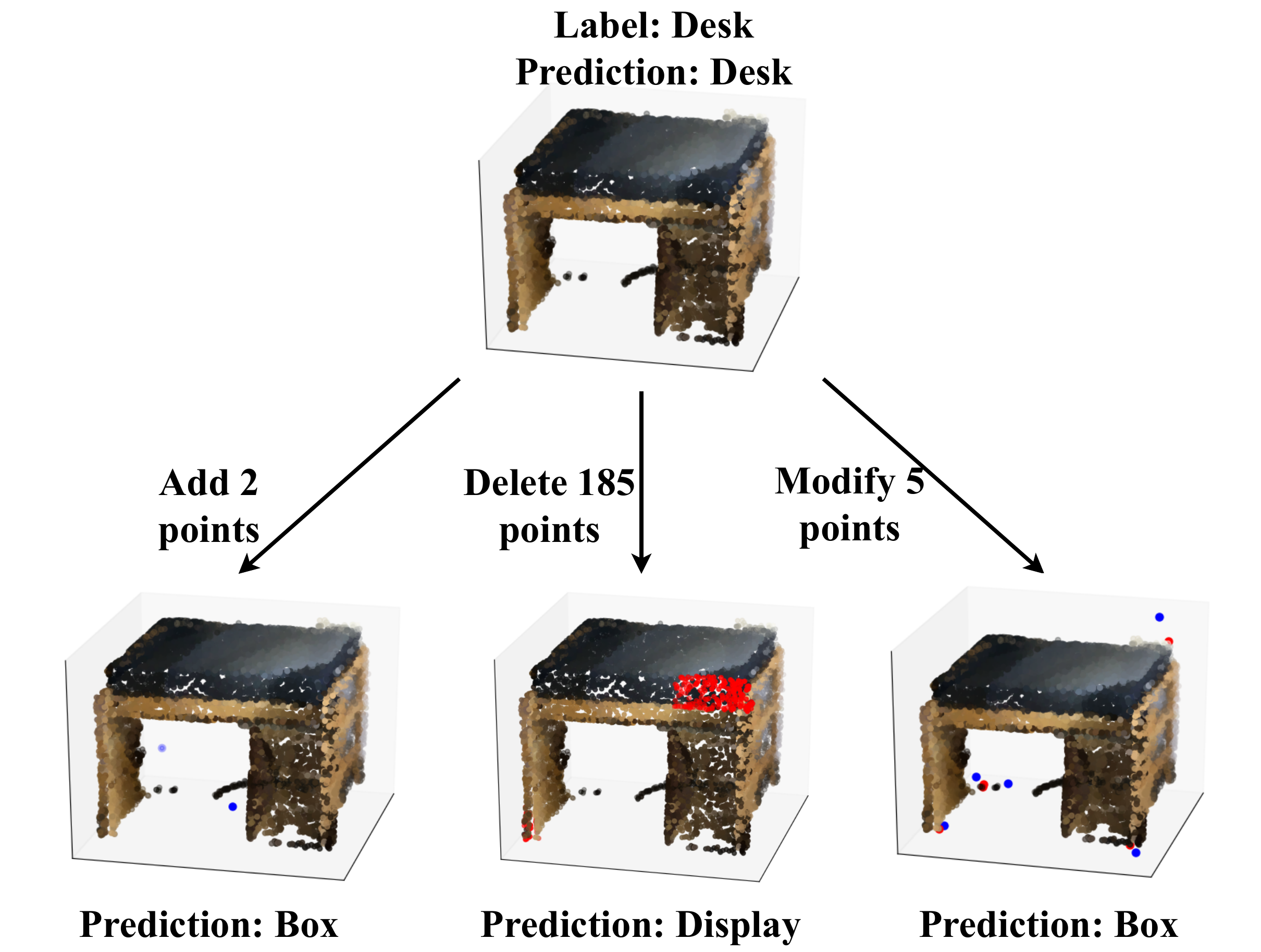}
    \caption{Illustration of \emph{point addition attack} (left), \emph{point deletion attack} (middle), and \emph{point modification attack} (right). Blue (or red) points are added (or deleted) by an attacker.}
    \label{fig:overview-three-attacks}
\end{figure}

\section{Details of Our Empirical Attacks.}
\label{app:attack}
In our experiments, we adopt the strong point addition, modification, and perturbation attacks developed by~\cite{xiang2019generating} and point deletion attack developed by~\cite{wicker2019robustness} to attack an undefended model. 
Roughly speaking, Xiang et al.~\cite{xiang2019generating} formulated point addition (or modification or perturbation) attack as an optimization problem, i.e., adversarial points can be crafted by minimizing a loss function using gradient descent. Wicker et al.~\cite{wicker2019robustness} developed an algorithm to identify a set of critical points in a point cloud whose removal would make a point cloud classifier predict an incorrect label for the point cloud.
    
Since there are no existing adversarial point cloud attacks tailored to randomized smoothing, PointGuard, and {\name}, we generalize existing attacks to them to compute Empirical Accuracy@$t$:

For \textbf{generalized point addition attack}, we iteratively add $t$ points using the attack in~\cite{xiang2019generating}. In particular, in the $i$th iteration ($i=1, 2, \cdots, t$), we  generate multiple noisy point clouds with Gaussian noise (or subsampled point clouds or sub-point clouds) from the testing point cloud with $i-1$ adversarially added points  in randomized smoothing (or PointGuard or {\name}). Then, we find a point such that the average loss (i.e., cross-entropy loss) of the base point cloud classifier on the noisy point clouds (or subampled point clouds or sub-point clouds) is maximized when the point is added to them. We use gradient descent to find the point. Moreover, to consider a powerful attack, we do not restrict the dimension values of the point. 

For \textbf{generalized point deletion attack}, we first generate multiple noisy point clouds (or subsampled point clouds or sub-point clouds) from a point cloud and then use the point deletion attack in~\cite{wicker2019robustness} to identify a set of critical points for each of them. Finally, we count the number of times for each point being identified as a critical point and delete the $t$ points with the largest counts from the point cloud.
    
Our \textbf{generalized point modification (or perturbation) attack} against randomized smoothing, PointGuard, and {\name} is a combination of our point addition and deletion attacks. Specifically, we first use our point deletion attack to delete $t$ points in a testing point cloud and then use our point addition attack to add $t$  points to the point cloud. 

\begin{figure}[t]
    \centering
    \includegraphics[width =0.9\linewidth]{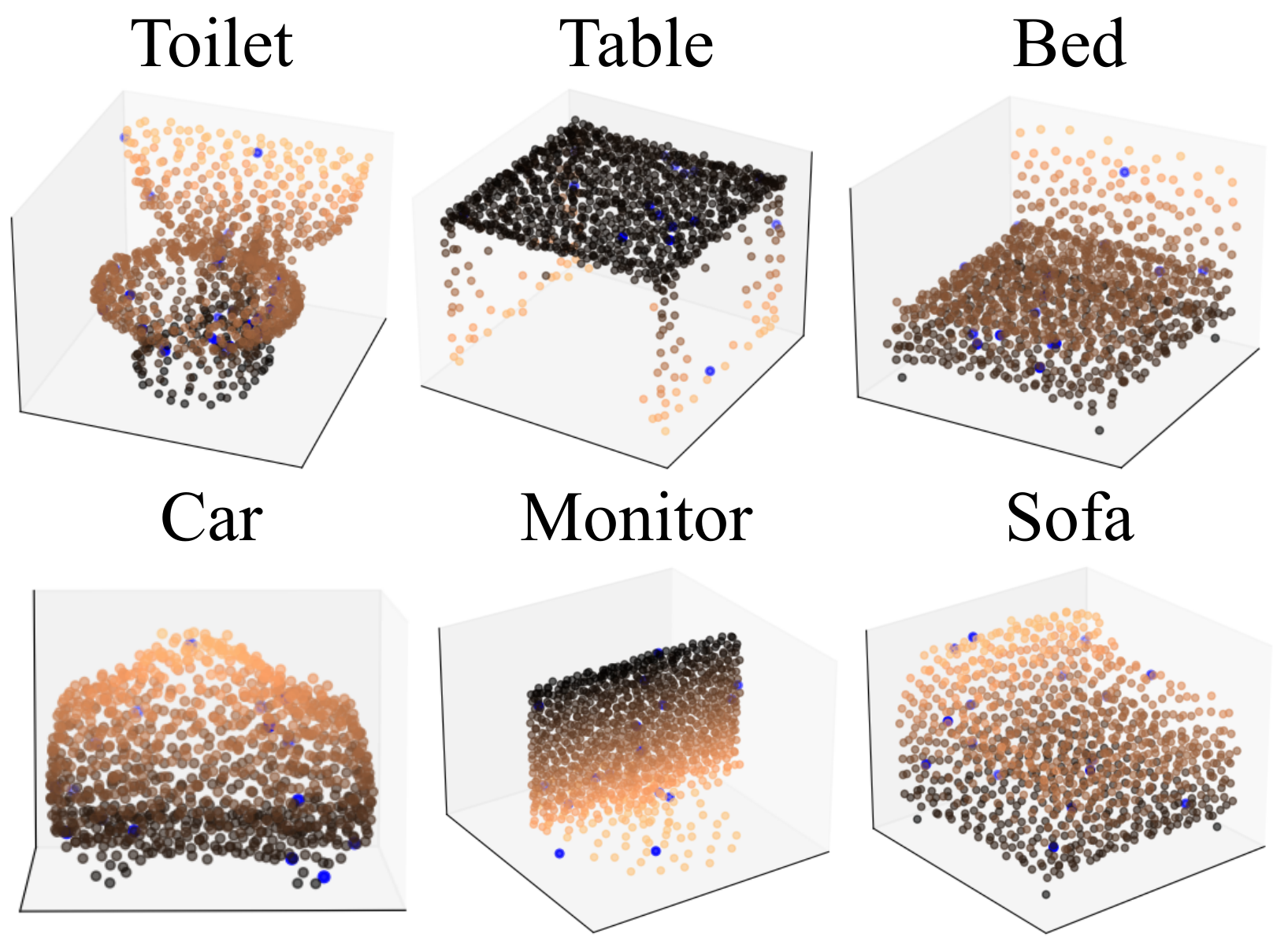}
    \caption{More visual illustrations of point completion with sub-point clouds. Each point cloud is reconstructed by a very small number of points (at most 32 points), colored in blue. Motivated by this observation, we utilize point cloud completion to build robust classifiers on the customer side in Scenario III.}
    \label{fig:teasar}
\end{figure}

\section{Additional Experiments}
\subsection{Comparing {\name} and PointGuard with Different $k$}
Certified defenses have accuracy-robustness trade-offs, which are controlled by their parameters (e.g., $m$ in PointCert and $k$ in PointGuard). By default, we use PointGuard with $k = 256$ such that PointGuard and PointCert have similar accuracy under no attacks for fair comparison of robustness. We also compare PointCert ($m=400$) and PointGuard with different $k$, where the same $k$ is used for both training and inference. Figure~\ref{fig:reb}(a) shows the results when PointGuard uses different $k$. When $k$ is very small, PointGuard can tolerate more perturbed points, but its certified accuracy under no attacks is much lower than PointCert. The reason is that PointGuard estimates probability bounds using a Monte-Carlo algorithm when computing certified robustness.

\subsection{Comparing {\name} with Deterministic PointGuard}
We can make PointGuard deterministic via fixing the seed in the random number generator. Moreover, we can extend our techniques for PointCert to derive tight certified robustness guarantees of such deterministic PointGuard. However, the certified accuracy of such deterministic PointGuard is low, as shown in Figure~\ref{fig:reb}(b). This is because a single newly added adversarial point can influence all subsampled point clouds in the worst-case for PointGuard. However, a single added adversarial point can only influence one sub-point cloud for PointCert because the sub-point clouds are disjoint. Note that, when the guarantees are probabilistic, PointGuard achieves larger certified accuracy because the probability that the worst-case happens is small and can be tolerated within the error probability.

\subsection{Comparing {\name} with Jia et al.~\cite{jia2022almost}} 
\begin{figure}
\centering
\subfloat[]{\includegraphics[width =0.24\textwidth]{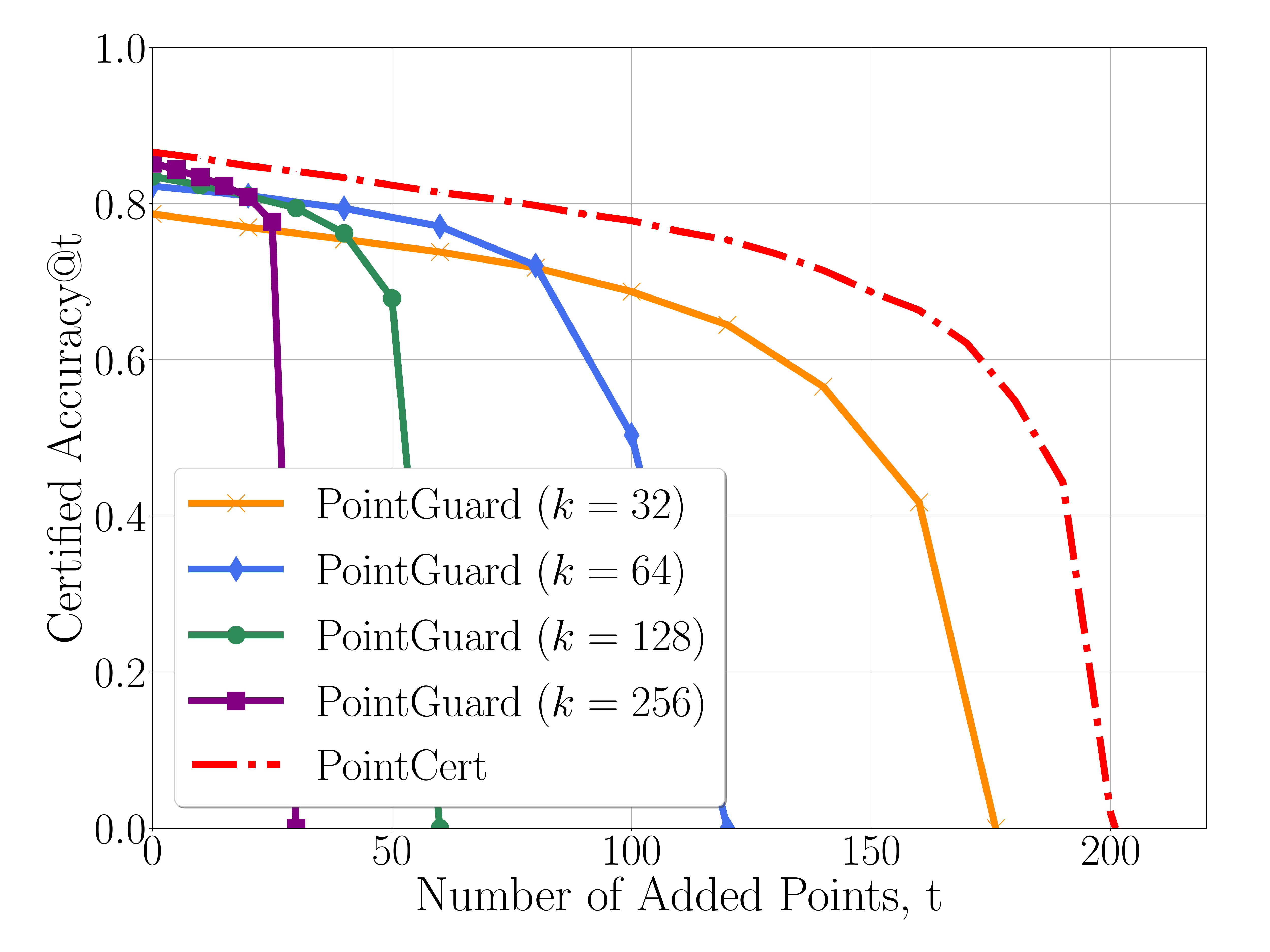}\label{rebuttal}}
\subfloat[]{\includegraphics[width =0.24\textwidth]{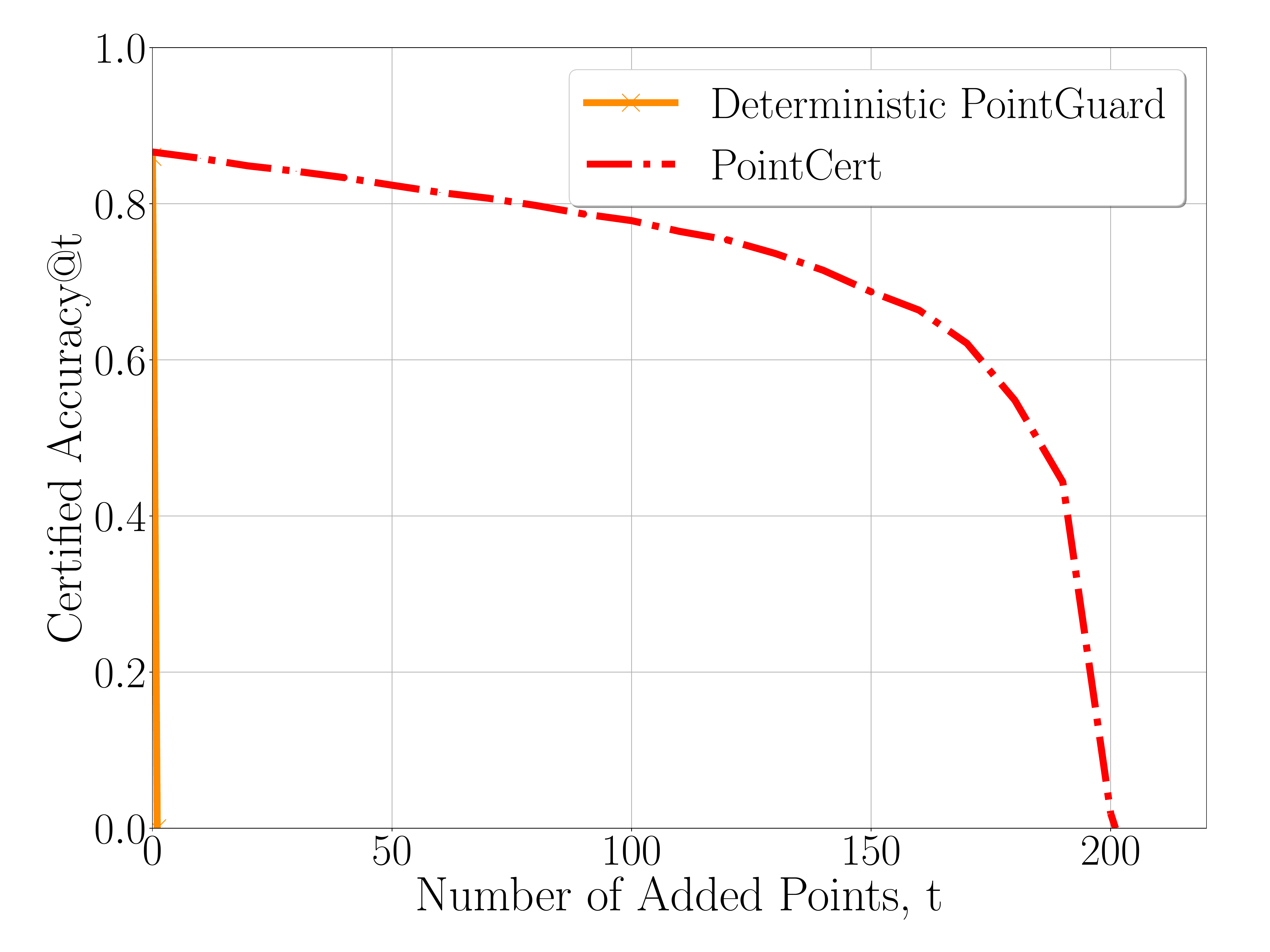}\label{rebuttal0}}
\\
\subfloat[]{\includegraphics[width =0.24\textwidth]{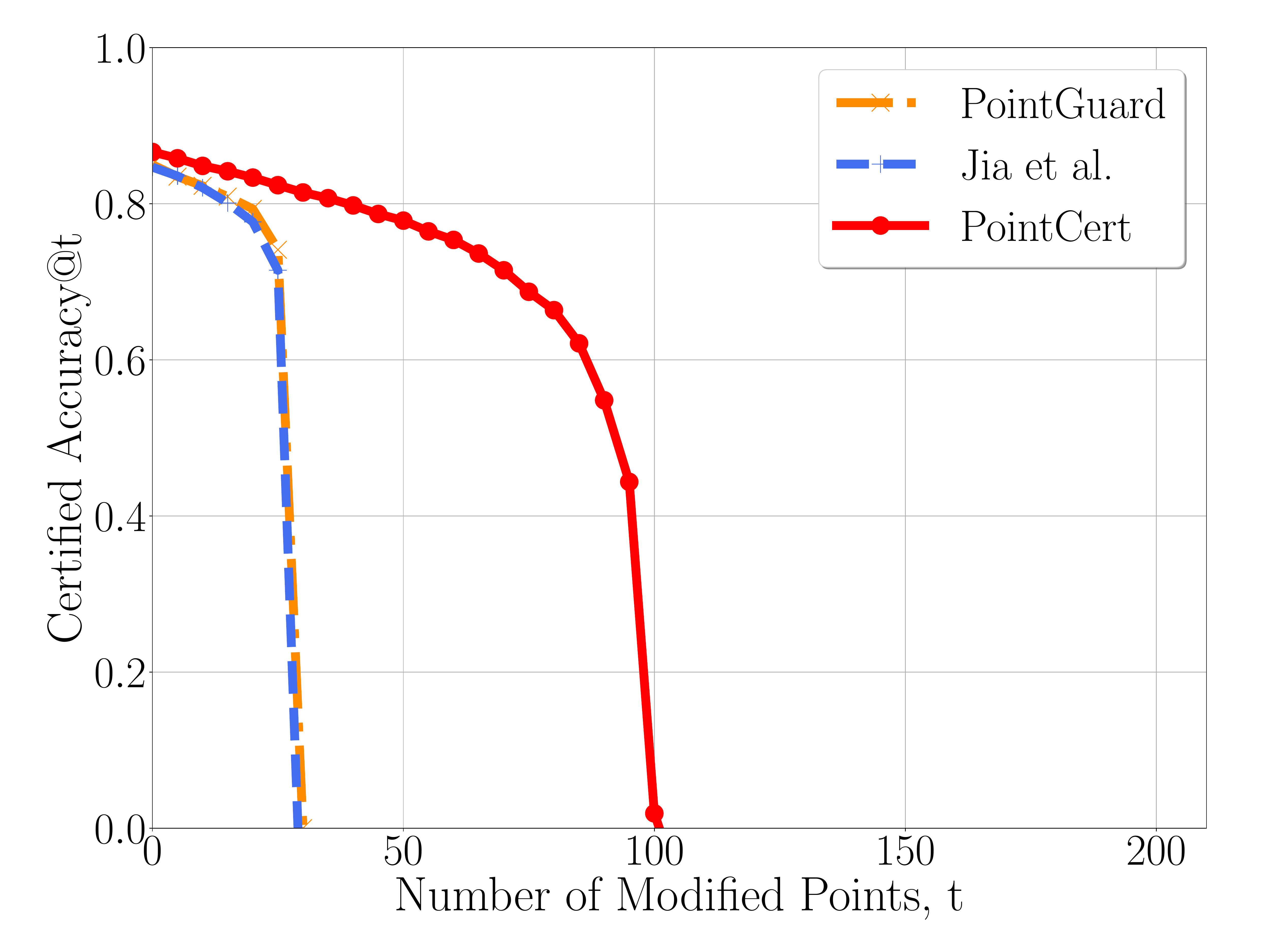}\label{rebuttal2}}
\caption{(a) Comparing {\name} and PointGuard with different $k$. (b) Comparing {\name} with deterministic PointGuard. (c) Comparing {\name} and Jia et al.~\cite{jia2022almost}.}
\label{fig:reb}
\end{figure}

Jia et al.~\cite{jia2022almost} develops almost tight $l_0$-norm certified robustness of top-$k$ predictions for image classification. Given a testing image $I$, their method creates different ablated inputs via retaining $b$ randomly selected pixels of $I$ and setting the remaining pixels to a special value. Then, they feed the ablated inputs to a base classifier and count the probabilities that the base classifier outputs for each label. In this way, they build a smoothed classifier that outputs  top-$k$ predictions with $l_0$-norm certified robustness. When extending their method to point cloud classification, we set $k=1$ and create ablated point clouds via retaining $b$ randomly selected points while setting the remaining points to a special value. From Figure~\ref{fig:reb}(c), we observe that Jia et al. achieves similar certified accuracy with PointGuard, which is lower than PointCert. The reason is that these two methods both use randomly selected/subsampled points for certification. We note that Jia et al. is only applicable to point modification attacks.

\begin{figure*}
    \centering
    \subfloat[Point addition/deletion attacks]{\includegraphics[width =0.5\textwidth]{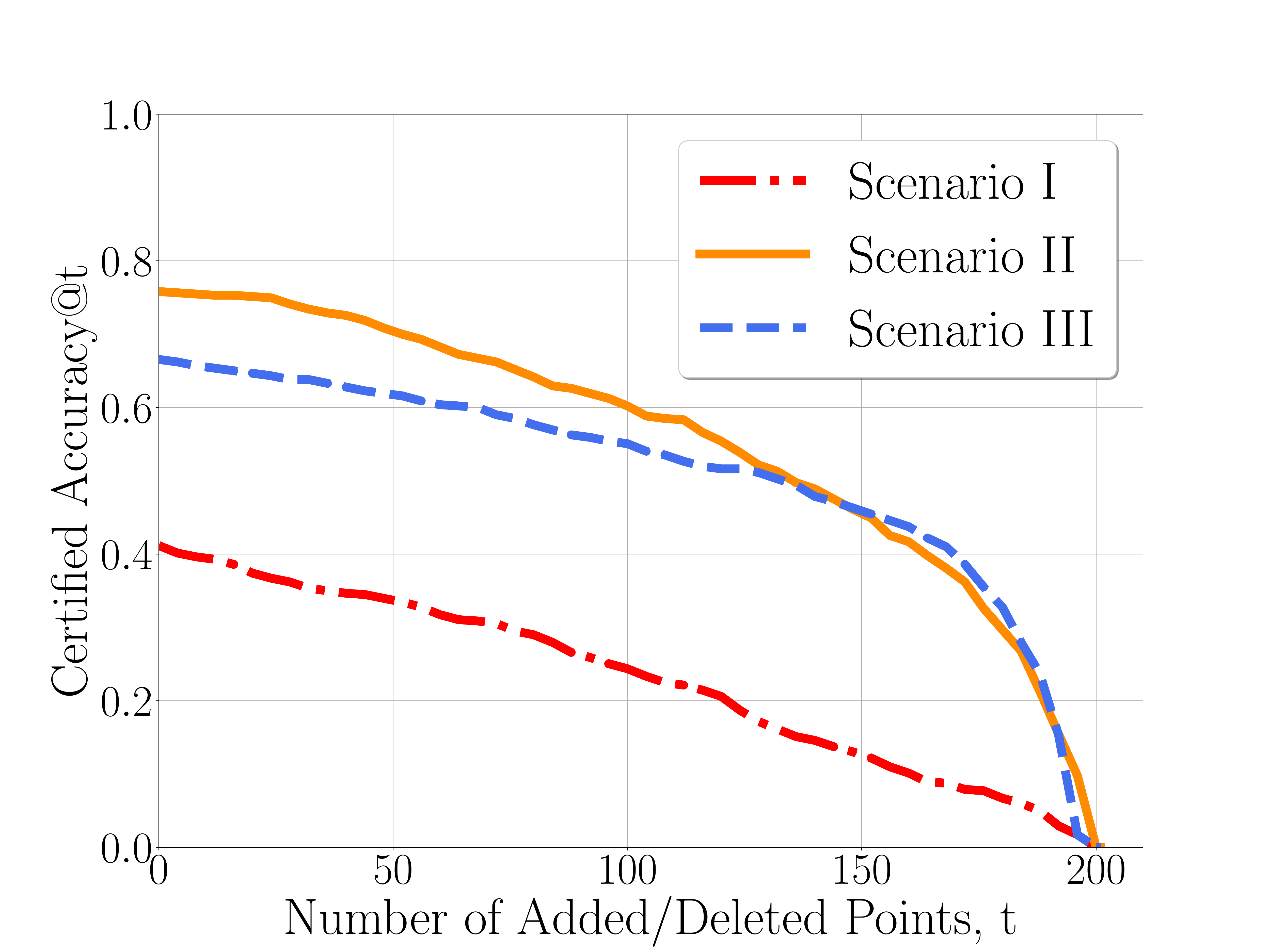}}    
    \subfloat[Point modification/perturbation attacks]{\includegraphics[width =0.5\textwidth]{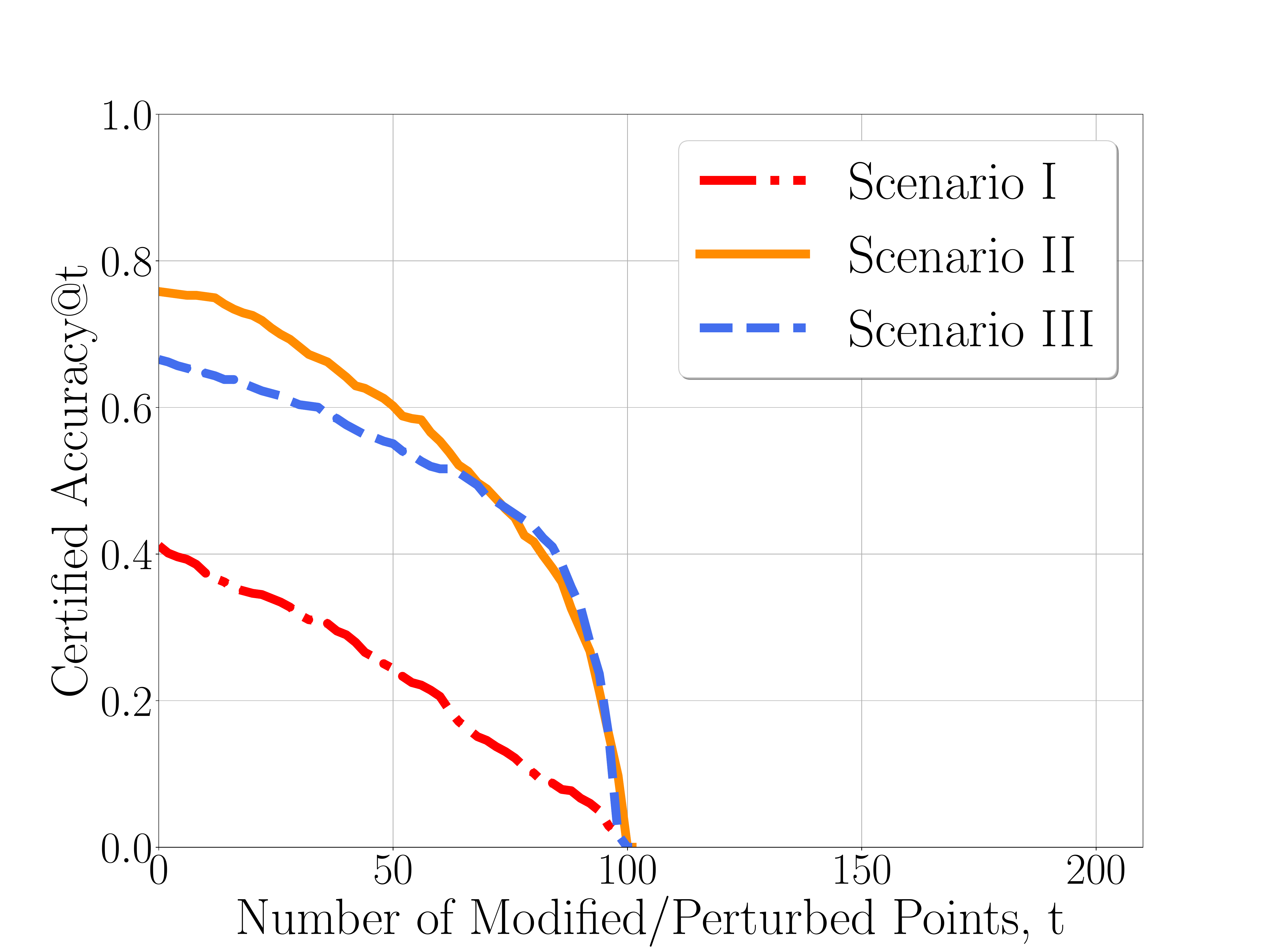}} 
    \caption{Comparing the certified accuracy of {\name} in the three application scenarios under different attacks. The dataset is ScanObjectNN-OBJ\_Only.}
    \label{fig:threescenarios-scanobjectnn_objonly}
\end{figure*}

\begin{figure*}
    \centering
    \subfloat[Point addition/deletion attacks]{\includegraphics[width =0.5\textwidth]{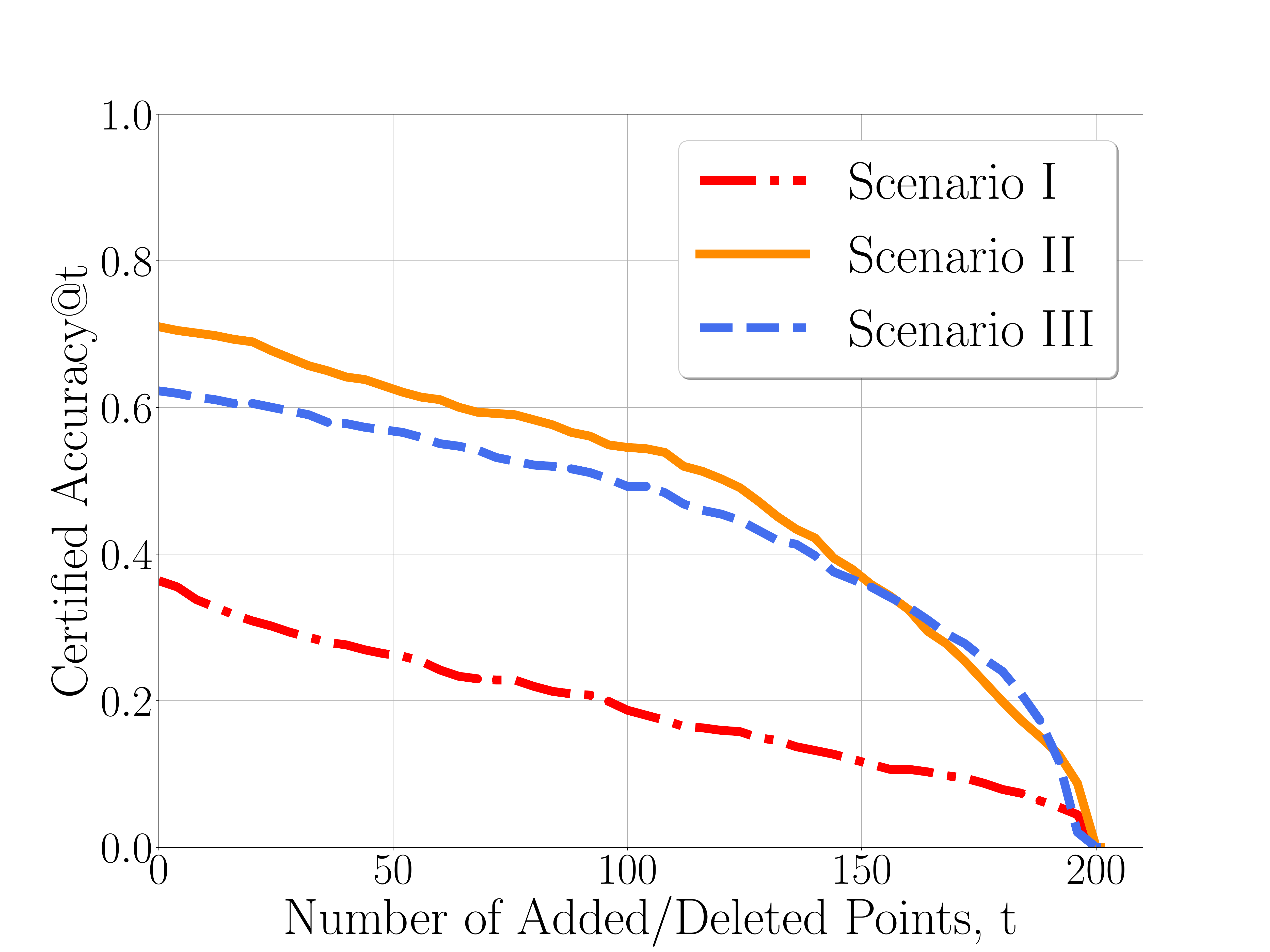}}    
    \subfloat[Point modification/perturbation attacks]{\includegraphics[width =0.5\textwidth]{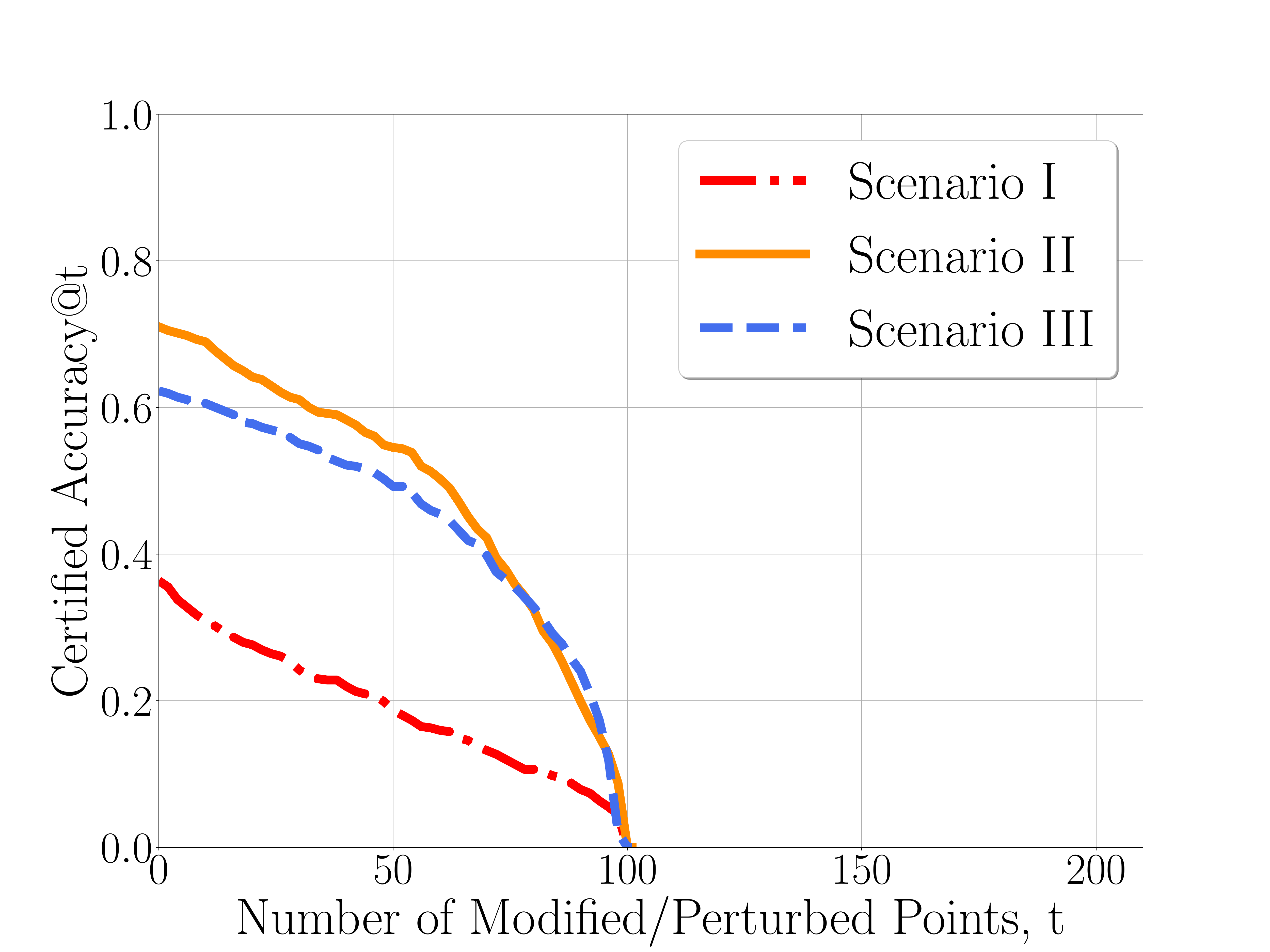}} 
    \caption{Comparing the certified accuracy of {\name} in the three application scenarios under different attacks. The dataset is ScanObjectNN-OBJ\_BG.}
    \label{fig:threescenarios-scanobjectnn_objbg}
\end{figure*}

\begin{figure*}
    \centering
    \subfloat[ScanObjectNN-OBJ\_Only]{\includegraphics[width =0.5\textwidth]{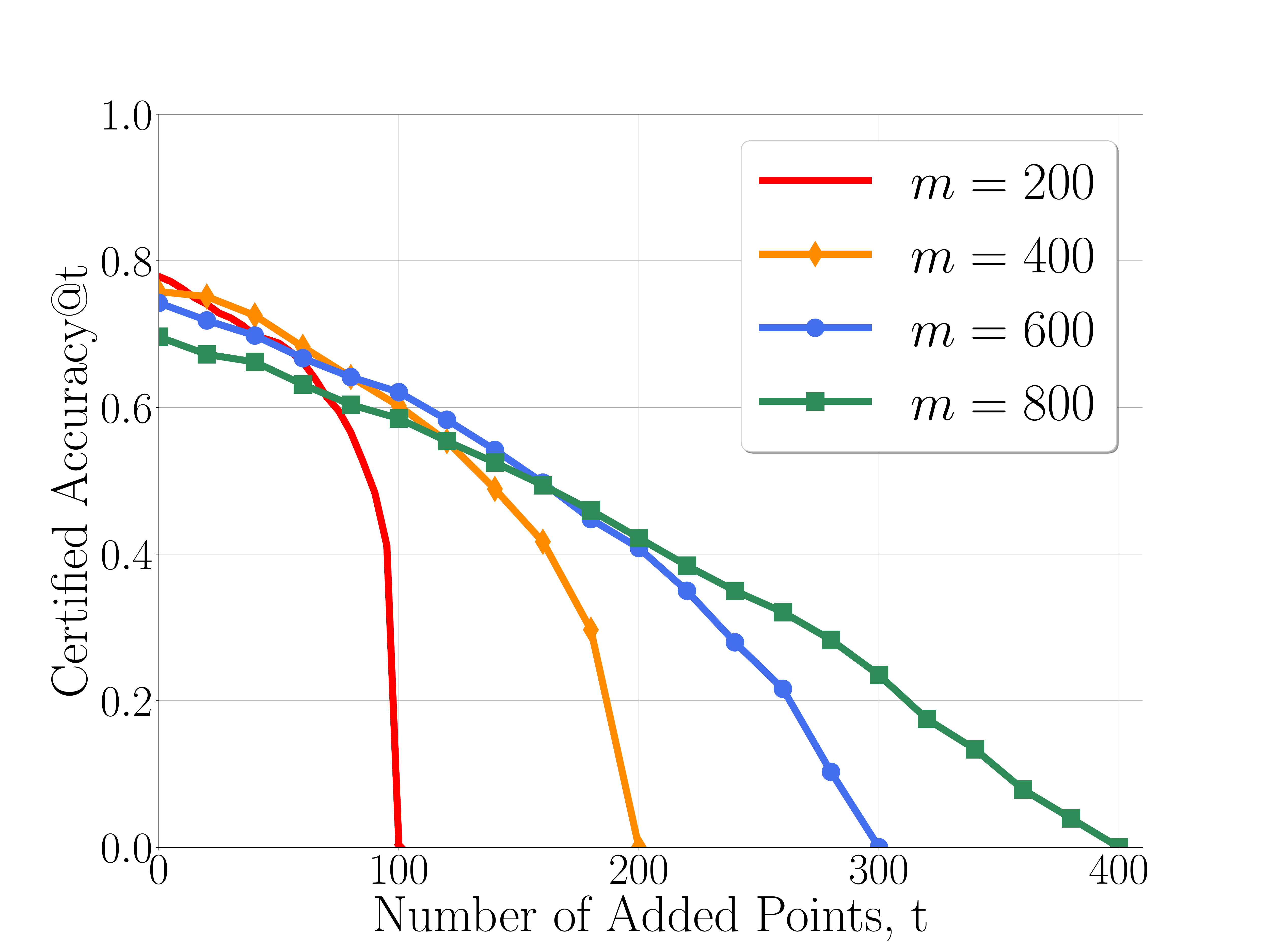}}
    \subfloat[ScanObjectNN-OBJ\_BG]{\includegraphics[width =0.5\textwidth]{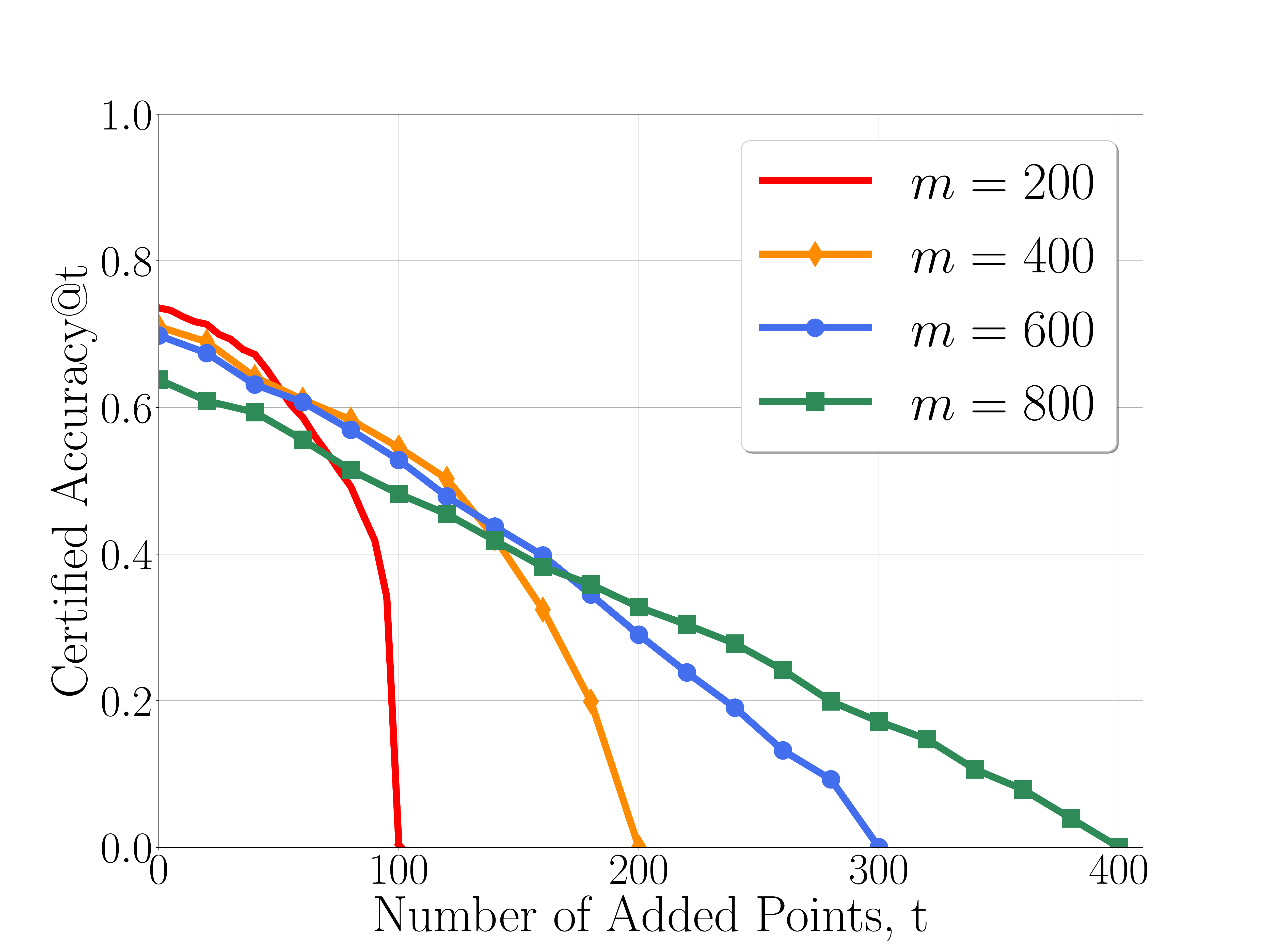}}
    \caption{Impact of $m$ on the certified accuracy of {\name} in Scenario II.}
    \label{fig:impact_of_n_ScanObjectNN}
\end{figure*}

\begin{figure*}
    \centering
    \subfloat[ScanObjectNN-OBJ\_Only]{\includegraphics[width =0.5\textwidth]{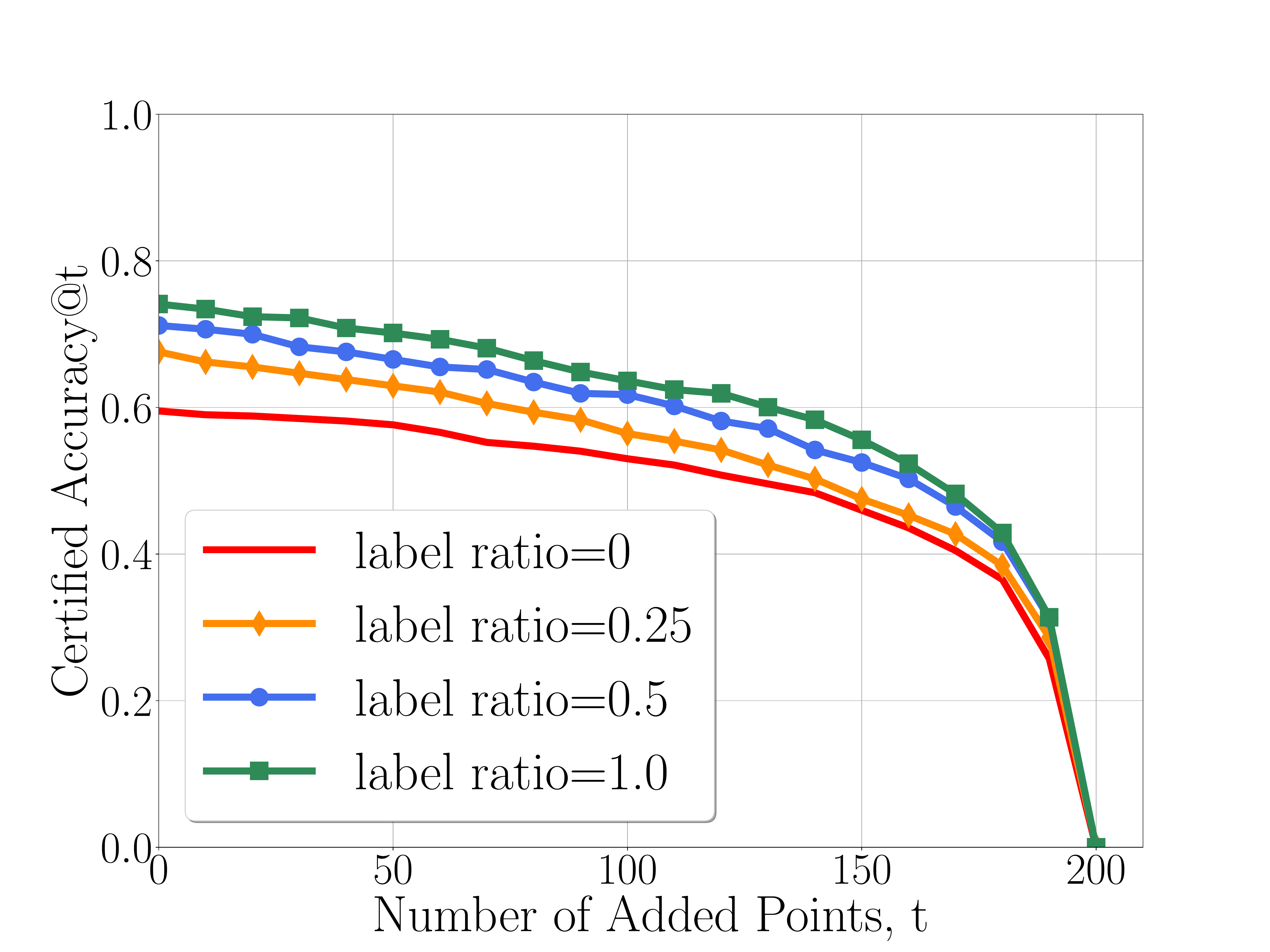}}
    \subfloat[ScanObjectNN-OBJ\_BG]{\includegraphics[width =0.5\textwidth]{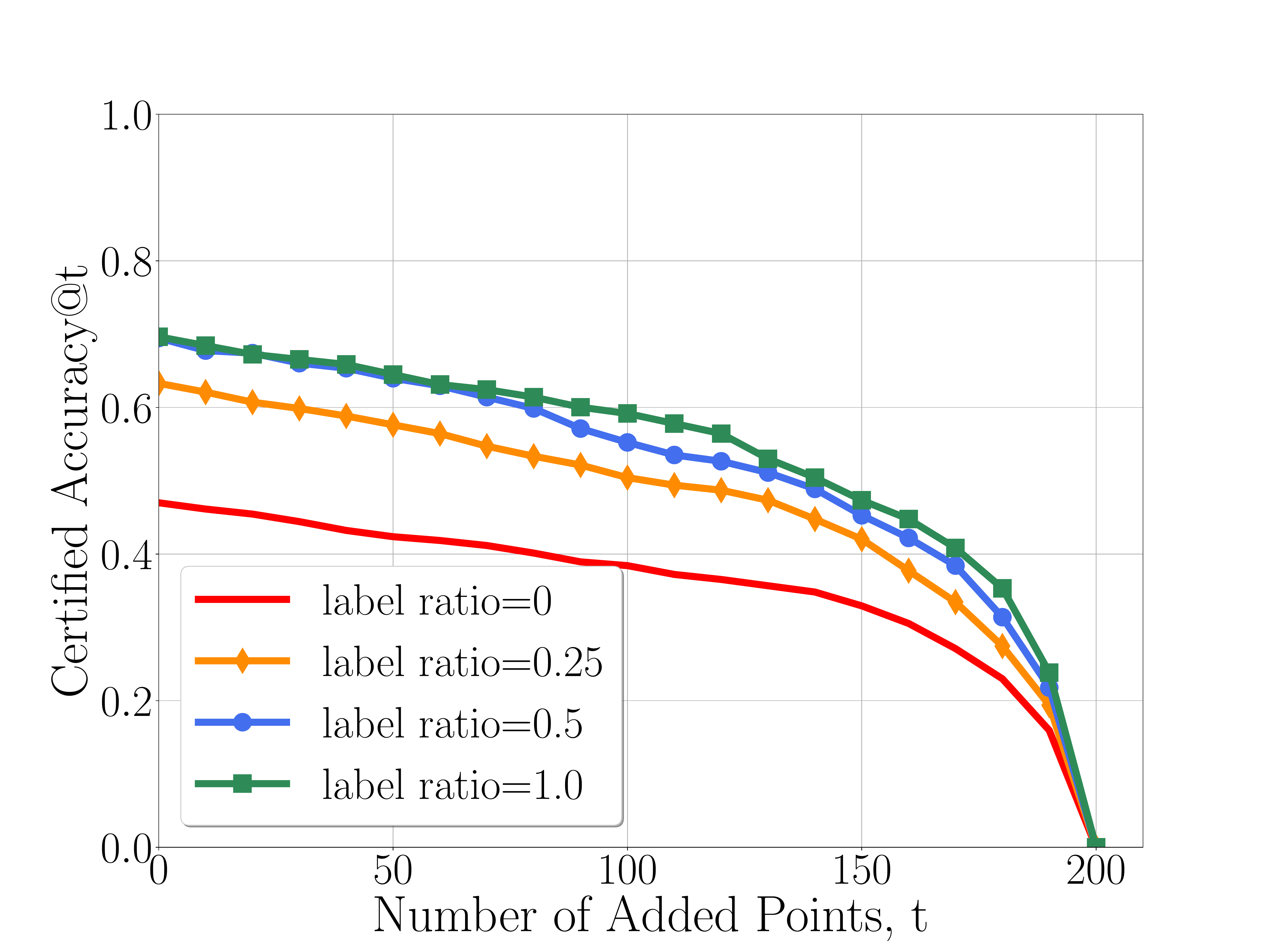}}
    \caption{Impact of the fraction of a customer's labeled point clouds on the certified accuracy of PointCert in Scenario III. }
    \label{fig:impact_of_label_ratio_ScanObjectNN}
\end{figure*}

\begin{figure*}
    \centering
    \includegraphics[width =\linewidth]{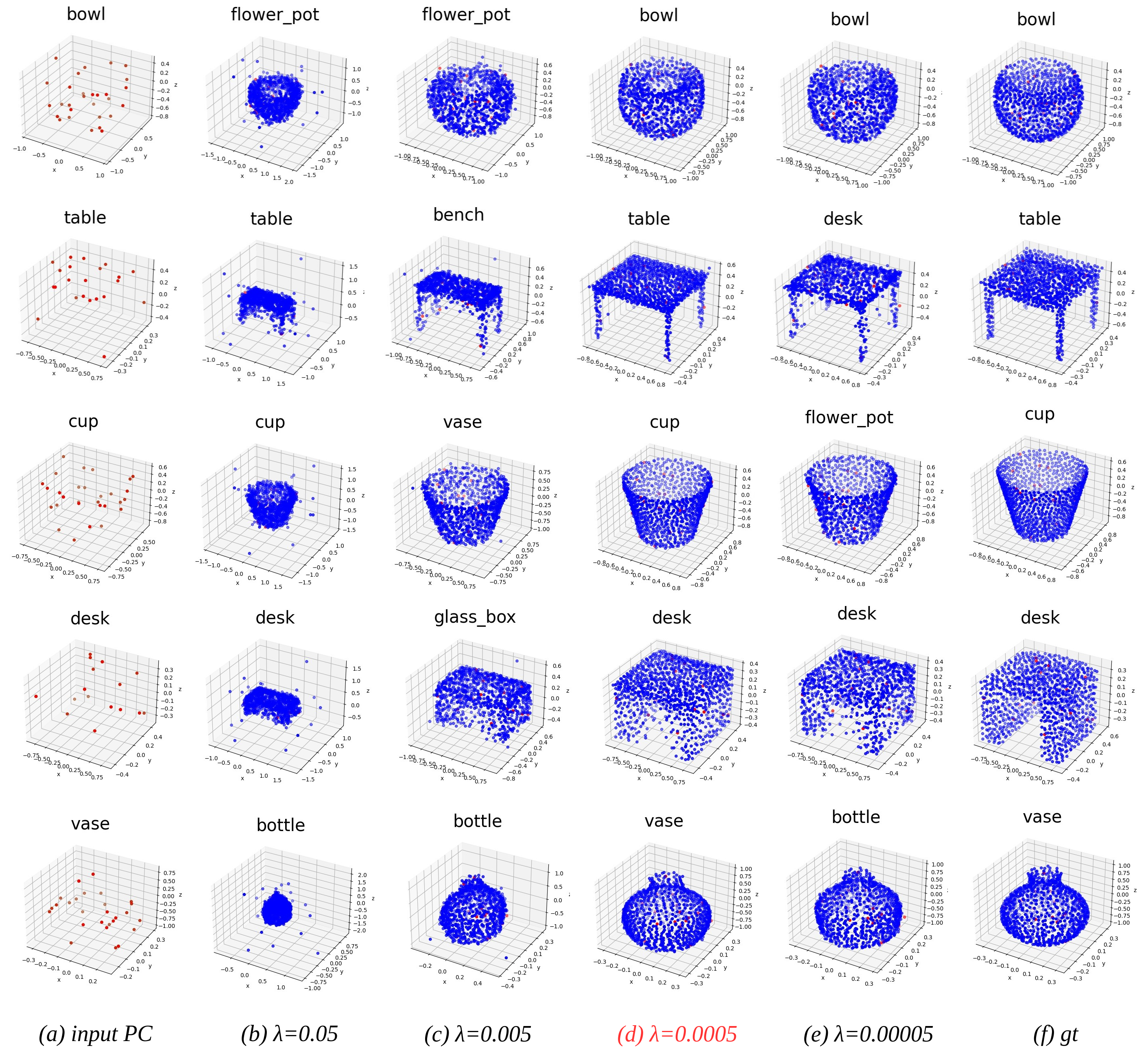}
    \caption{Examples of completed point clouds outputted by PCN when $\lambda$ varies. PC stands for point cloud and gt stands for ground truth. The word above a point cloud is its label predicted by the base point cloud classifier. The dataset is ModelNet40. The highlighted $\lambda=0.0005$ is used in our experiments.}
    \label{fig:lambda_appendix}
\end{figure*}
\end{document}